\documentclass[12pt]{article}
\usepackage{amsmath,amsthm,amscd,amssymb,array}

\begin{document}

\hfill NTZ 04/2001
\vspace*{1cm}

\begin{center}
{\Large{\bf Algebraic renormalization
\\
\medskip
of twisted N = 2 supersymmetry
\\
\medskip\smallskip
with Z = 2 central extension }}
\\
\vspace*{2cm}
{\large{\sc B. Geyer}}\footnote{e-mail: geyer@itp.uni-leipzig.de}
\\
\smallskip
{\it Universit\"at Leipzig, Naturwissenschaftlich-Theoretisches Zentrum}
\\
{\it and Institut f\"ur Theoretische Physik, D--04109 Leipzig, Germany}
\\
\bigskip
{\large{\sc D. M\"ulsch}}\footnote{e-mail: muelsch@informatik.uni-leipzig.de}
\\
\smallskip
{\it Wissenschaftszentrum Leipzig e.V., D--04103 Leipzig, Germany}
\\
\vspace*{2cm}
{\small{\bf Abstract}}
\\
\end{center}

\begin{quotation}
\noindent {\small
{We study the renormalizability of (massive) topological QCD based on the 
algebraic BRST technique by adopting a non--covariant Landau type gauge 
and making use of the full topological superalgebra. The most general 
local counter terms are determined and it is shown that in the presence 
of central charges the BRST cohomology remains trivial. By imposing an 
additional set of stability constraints it is proven that the matter 
action of topological QCD is perturbatively finite.}}
\end{quotation}

\vspace*{.5cm}
\noindent

PACS: 12.60.J, 03.70, 11.10.G 
\\
PACS: Topological QCD; $N = 2$ twisted supersymmetry; dimensional reduction;

\bigskip\medskip
\begin{flushleft}
{\large{\bf 1. Introduction}}
\end{flushleft}
\bigskip
Topological quantum field theories (TQFT) are characterized by observables 
depending only on the topology of the manifold on which these theories 
are defined \cite{1}. The classical example is topological Yang--Mills 
theory (TYM) on a four--manifold as proposed by Witten \cite{2} 
whose correlation functions, if computed in the weakly coupled ultraviolet 
limit, turn out to be related to the Donaldson invariants \cite{3}. 
On the other hand, the Donaldson invariants of smooth four--manifolds are 
closely related to $N = 2$ supersymmetric Yang--Mills theories (SYM). 
In fact, by restricting to flat Euclidean space--time TYM can be 
reformulated through a twisting of $N = 2$ SYM in Wess-Zumino 
gauge \cite{2}. The resulting theory has a fermionic symmetry being 
identified with the topological shift operator $Q$ of the twisted $N = 2$ 
SUSY algebra. In the cohomological formulation of Labastida--Pernici \cite{4} 
and Baulieu--Singer \cite{5} TYM is interpreted as gauge--fixed action of 
the Pontryagin term and $Q$, after identifying the $R$--charge with the ghost 
number, is interpreted as the BRST operator of the equivariant part
of TYM. As a result of that 
identification the BRST cohomology becomes completely trivial. In the 
derivation of ultraviolet finiteness properties of various topological 
models and in the construction of their observables, besides $Q$, still 
another fermionic symmetry, the so--called vector supersymmetry, plays an 
important role being generated by the vector charge $\bar{Q}_\mu$ of the  
twisted $N = 2$ SUSY algebra \cite{6,7,8,9,10}. 

Seiberg and Witten have studied also the strongly coupled infrared limit
of twisted $N = 2$ SYM which leads to a 
new class of four--manifold invariants, the so--called Seiberg--Witten 
invariants \cite{12}. The Seiberg--Witten theory can be considered as the 
twisted version of $N = 2$ supersymmetric Maxwell theory with one 
massless hypermultiplet \cite{13}. In a second paper, Seiberg and Witten
have studied the strong coupled infrared limit of topological quantum
chromodynamic (TQCD), being the twisted version of $N = 2$ SYM coupled 
to a standard hypermultiplet in the fundamental representation \cite{14}. 
In Ref.~\cite{15} it has been shown that by introducing the $N = 2$
supersymmetric bare mass term to the hypermultiplet the resulting massive
TQCD interpolates between the Donaldson and the Seiberg--Witten 
theory. Such hypermultiplets have a non--vanishing central charge and, 
therefore, an extension of the $N = 2$ SUSY algebra is required. 
A useful approach to understand the geometry of the TQCD
is based on the Mathai--Quillen formalism \cite{16}.

In the present paper we investigate the renormalization properties of 
twisted $N = 2$ SUSY with two central extensions. In extending the 
analysis of Ref.~\cite{17} we study the ultraviolet behaviour of TQCD 
in the framework of algebraic renormalization \cite{18,18a} by exploiting both 
the topological shift symmetry $Q$ and the vector supersymmetry $\bar{Q}_\mu$.

The renormalization properties of twisted $N = 2$ SUSY {\rm without}
central extensions have been widely investigated for particular gauges
\cite{6,7,17,19,20,21}. 
In Ref. \cite{20} it has been shown that the cohomological 
nature of Witten's topological model is totally insensitive to quantum 
corrections and, in accord with a previous result \cite{5}, that gauge 
anomalies are absent to all orders of perturbation theory. 
This analysis has been 
extended in Ref. \cite{17} by choosing a non--covariant Landau type gauge. 
This gauge has the merit that the associated vector supersymmetry 
$\bar{Q}_\mu$ is linearly realized and, therefore, can be employed as 
a stability constraint in order to improve the ultraviolet finiteness 
properties of that model. 

Here, we study the renormalizability of a slightly more general
extension of TQCD which is formed by twisting $N = 2$ SUSY with 
$Z = 2$ central charges. Let us recall, that for $N = 2$ the twisting 
procedure is unique. 
An undesired feature which is displayed by twisting $N = 2$ SYM 
coupled to a standard massive hypermultiplet \cite{14} 
is the fact that it fails to be invariant under $R$--symmetry, the 
latter being broken into $Z_2$. In order to circumvent this problem we 
consider TQCD in the presence of two central charges $Z$ and $\bar{Z}$, 
being complex conjugated to each other. It can be shown that this 
topological model preserves $R$--symmetry and, therefore, the ghost number
conservation if one formally ascribes to $Z$ and $\bar{Z}$ the $R$--weights
$R(Z) = 2 R(Q)$ and $R(\bar{Z}) = 2 R(\bar{Q}_\mu)$, respectively.

Furthermore, by adopting the non--covariant Landau type gauge it will be 
proven that the whole set of stability constraints considered in 
Ref.~\cite{17} can be applied also in that general case -- 
except for the so--called ghost for the ghost equation. Moreover, we shall 
be able to give an additional set of constraints which ensures the 
perturbative finiteness of the matter part of TQCD, i.e., the twisted 
hypermultiplet is not subjected to any renormalization. It is worthwhile 
to emphasize here that the algebraic proof of that finiteness property 
extends to all orders of perturbation theory and does not rely on the 
existence of any regularization scheme.

The paper is organized as follows. Sect.~2 reviews the main 
features of TQCD. Its purpose is, at first, to introduce our notations and 
to make this paper reasonably self--contained and, secondly, to bring
separately into the play the various elements of $N = 2$ TQCD with two
central charges thereby also motivating the choice of (non--covariant)
Landau type gauge. 
In Subsect.~2.1 we introduce the so-called equivariant part of the action 
of TYM by fixing the topological shift symmetry in an covariant 
$\xi$-gauge. This procedure closely parallels the construction,
in a Feynman type gauge $(\xi = 1)$, of the topological action in 
Ref.~\cite{21} but, in contrast to it, we introduce a parameter $\xi$ into 
the twisted $N = 2$ SUSY algebra for being able also to select the Landau 
type gauge $(\xi = 0)$.
In Subsect.~2.2 we extend the topological Yang-Mills theory by coupling it
to a massive hypermultiplet (and its hermitian conjugate) thus obtaining
the equivariant part of the action of TQCD. This is achieved by a 
non--trivial dimensional reduction of a $D = 6$ dimensional $N = 1$ SYM 
containing a gauge multiplet and a massless hypermultiplet in the 
adjoint and some (e.g., the fundamental) representation of the gauge algebra 
$Lie(G)$, respectively. The main body of that derivation is postponed to
Appendix B.
In Subsect.~2.3 we introduce the complete action of TQCD by fixing also the 
remaining gauge symmetry. It turns out that only in the Landau 
type gauge, $\xi = 0$, the complete action is invariant under both the 
topological shift symmetry $Q$ and the vector supersymmetry $\bar{Q}_\mu$. 
In Sect.~3 we construct the action of TQCD in a non--covariant Landau 
type gauge which allows to realize the vector supersymmetry linearly and,
lateron, to employ it as an additional stability constraint. 
In Sect.~4 we enclose all the symmetry operators of the theory into a 
single nilpotent BRST operator $s_T$ by associating to each generator of the 
topological superalgebra a global ghost. We establish the corresponding 
classical Slavnov-Taylor identity and derive the gauge conditions, 
the antighost equations and some global constraints, being related to the 
non--covariant Landau type gauge. 
In Sect.~5 the problem of renormalizability is treated by standard 
cohomological methods. It is proven that the BRST cohomology is trivial 
and the most general local counterterms are determined.
Appendix A contains the Euclidean spinor conventions being used 
thourough this paper.
Appendix B gives a detailed derivation of the twisted $N = 2$ SYM coupled 
to a (massive) hypermultiplet with two central charges by using the method 
of dimensional reduction \cite{22}. 

In this paper we use the convention $[ T^i, T^j ] = f^{ijk} T^k$ and 
${\rm tr}(T^i T^j) = \delta^{ij}$, the antihermitean generators $T^i$ being 
a basis of the Lie algebra $Lie(G)$ of the gauge group $G$, which we 
assume to be a simple compact Lie group, in some representation $\cal R$,
e.g., a fundamental representation. We also adopt the matrix 
notation $\varphi = \varphi^i T^i$ and 
$\delta/\delta \varphi = T^i \delta/\delta \varphi^i$ for any field
$\varphi^i$ transforming according to the adjoint representation of $G$. 
--- Furthermore, we use the convention $\bar{\varphi}$
for the hermitian conjugate of some field $\varphi$.


\newpage
\begin{flushleft}
{\large{\bf 2. Topological QCD with Z = 2 central extensions}}
\end{flushleft}
\bigskip
\begin{flushleft}
{2.1. {\it Topological Yang--Mills theory: Equivariant part}}
\end{flushleft}
\bigskip
One of the possibilities to introduce TYM consists in twisting a set of 
conventional (spinorial) supercharges $Q_A^{\!~~a}$ and 
$\bar{Q}_{\dot{A} a}$ of $N = 2$ SYM in 
$D = 4$ dimensional Euclidean space--time \cite{2}. Thereby, the $N = 2$ 
gauge multiplet consists of a (antihermitean) gauge field $A_\mu$, the 
$Sp(2)$--doublets of chiral and of anti-chiral 2--spinors, 
$\lambda_A^{\!~~a}$ and $\bar{\lambda}_{\dot{A} a}$, respectively, and a 
complex scalar field $\phi$. In order to close the SUSY algebra it 
is necessary to introduce a (symmetric) auxiliary field 
$\chi_{ab} = \chi_{ba}$. All the fields of that off--shell gauge multiplet
$V = \{A_\mu, \lambda_A^{\!~~a}, \bar{\lambda}_{\dot{A} a}, \phi, \bar{\phi},
\chi_{ab} \}$ are in the adjoint 
representation and take their values in the Lie algebra $Lie(G)$ of some 
compact gauge group $G$. The rotation group of Euclidean space--time, 
$SO(4)$, is locally isomorphic to $SU(2)_L \otimes SU(2)_R$ and 
the spinor indices in the fundamental representation of $SU(2)_L$ and 
$SU(2)_R$ will be denoted by $A = 1, 2$ and $\dot{A} = \dot{1}, \dot{2}$, 
respectively. The global internal symmetry group of $N = 2$ SUSY is 
$Sp(2) \otimes U(1)_R$ corresponding to symplectic rotations and chiral 
transformations ($R$--symmetry). The $R$--charges (chiral weights) of 
$Q_A^{\!~~a}$ and $\bar{Q}_{\dot{A} a}$ are $1$ and $-1$, respectively. 
The internal $Sp(2)$ indices, labelling the different $N = 2$ charges, are 
denoted by $a = 1,2$. 

In the absence of a central extension, i.e., in a theory without massive
fields which will be considered first, the $N = 2$ SUSY algebra in 
the Wess--Zumino gauge is characterized by the eight spinorial supercharges
$Q_A^{\!~~a}$ and $\bar{Q}_{\dot{A} a}$ which, together with the generator 
$P_\mu \equiv i \partial_\mu$ of space--time translations, obey the relations 
\begin{align}
\label{2.1}
\{ Q_A^{\!~~a}, Q_B^{\!~~b} \} &= 
- 4 \epsilon^{ab} \epsilon_{AB} \delta_G(\phi),
\nonumber
\\
\{ Q_A^{\!~~a}, \bar{Q}_{\dot{B} b} \} &=  
- 2 \delta^a_b (\sigma^\mu)_{A \dot{B}} ( P_\mu + i \delta_G(A_\mu) ),
\\
\{ \bar{Q}_{\dot{A} a}, \bar{Q}_{\dot{B} b} \} &= 
\phantom{-}
4 \epsilon_{ab} \epsilon_{\dot{A} \dot{B}} \delta_G(\bar{\phi}),
\nonumber
\end{align}
with $\bar{\phi}$ being the hermitean conjugate of $\phi$
(The conventions of Euclidean spinor algebra are collected in Appendix A;
the on--shell version of the algebra (\ref{2.1}) is derived in Appendix B). 
Since in Wess--Zumino gauge the supersymmetry is realized nonlinearly the 
algebra (\ref{2.1}) closes only modulo the field dependent gauge 
transformations $\delta_G(\omega)$, $
\omega = \{ A_\mu, \phi, \bar{\phi} \}$, respectively (cf., Eqs.~(B.2)). 

As explained in 
Ref.~\cite{2}, TYM is obtained from $N = 2$ SYM by replacing the group
$SU(2)_L \otimes Sp(2)$ through its diagonal subgroup or, in other words, by 
identifying the internal index $a$ with the spinor index $A$. According to
this twisting procedure one constructs from the generators
$Q_A^{\!~~a}$ and $\bar{Q}_{\dot{A} a}$  the twisted ones, 
\begin{equation}
\label{N2S}
Q = \hbox{\large$\frac{1}{2}$} \epsilon^{AB} Q_{AB},
\qquad
\bar{Q}_\mu = \hbox{\large$\frac{1}{2}$}
i (\sigma_\mu)^{\dot{A} B} \bar{Q}_{\dot{A} B},
\qquad
Q_{\mu\nu} = \hbox{\large$\frac{1}{2}$} (\sigma_{\mu\nu})^{AB} Q_{AB},
\end{equation}
being a scalar $Q$, a vector $\bar{Q}_\mu$ and a self--dual tensor
$Q_{\mu\nu}$, respectively. Since it turns out that TYM is already 
completely specified by the topological shift symmetry $Q$ and the vector 
supersymmetry $\bar{Q}_\mu$, we actually do not take into account the 
self--dual tensor supersymmetry $Q_{\mu\nu}$ in the following. 
In terms of $Q$ and $\bar{Q}_\mu$, from (\ref{2.1}) for the twisted (or 
topological) superalgebra we get \cite{8,9} 
\begin{equation}
\label{2.2}
\{ Q, Q \} = - 2 \delta_G(\phi),
\qquad
\{ Q, \bar{Q}_\mu \} = - i P_\mu + \delta_G(A_\mu),
\qquad
\{ \bar{Q}_\mu, \bar{Q}_\nu \} = - 2 \delta_{\mu\nu} \delta_G(\bar{\phi}).
\end{equation}

From the first of these relations it follows that scalar supercharge $Q$ is 
equivariantly nilpotent, i.e. it squares to gauge transformations 
$\delta_G(\phi)$ generated by $\phi$, and, applying the Jacobi identity on 
$\phi$, it follows that the field $\phi$, from which the Donaldson invariant 
is constructed, must be $Q$--invariant, $Q \phi = 0$. 

The second relation, being typical for a topological theory, states that, 
due to the existence of the vector supercharge $\bar{Q}_\mu$, the space--time 
translations $P_\mu$ can be represented as a $Q$--anticommutator modulo 
the gauge transformation $\delta_G(A_\mu)$. This allows, starting from the
Donaldson invariant, $\cal O$, to construct all the (global) observables of 
TYM \cite{2} by applying successively the vector supercharge 
$\bar{Q}_\mu$ \cite{21} (see also [1(d)]), 
\begin{equation}
\label{2.3}
{\cal O} \equiv
\int d^4x\, {\rm tr}\, \phi^2, 
\qquad
\bar{Q}_\mu {\cal O},
\qquad
\bar{Q}_\mu \bar{Q}_\nu {\cal O},
\qquad
\bar{Q}_\mu \bar{Q}_\nu \bar{Q}_\rho {\cal O},
\qquad
\bar{Q}_\mu \bar{Q}_\nu \bar{Q}_\rho \bar{Q}_\sigma {\cal O}.
\end{equation}
(It is worthwhile to note that the same observables can also be recovered 
through the so--called equivariant cohomology, compare, e.g.,~Ref.~\cite{23}). 
In this manner one obtains that part of the action of TYM which results 
by fixing, in a Feynman type gauge, the topological shift symmetry \cite{21}:
\begin{equation}
\label{2.4}
W_{\rm T}^{(\xi = 1)} = \frac{1}{4} \int d^4x\, {\rm tr}\,
F_{\mu\nu} \tilde{F}^{\mu\nu} - \frac{1}{24} \epsilon^{\mu\nu\rho\sigma} 
\bar{Q}_\mu \bar{Q}_\nu \bar{Q}_\rho \bar{Q}_\sigma 
\int d^4x\, {\rm tr}\, \phi^2, 
\end{equation}
where
\begin{gather*}
\tilde{F}_{\mu\nu} = \hbox{\large$\frac{1}{2}$} 
\epsilon_{\mu\nu\rho\sigma} F^{\rho\sigma},
\qquad
F_{\mu\nu} = \partial_\mu A_\nu - \partial_\nu A_\mu + [ A_\mu, A_\nu ],
\qquad
D_\mu = \partial_\mu + [ A_\mu, ~\cdot~ ];
\end{gather*}
here, $\tilde{F}_{\mu\nu}$ is the dual of the YM field strenght
$F_{\mu\nu}$ and $D_\mu$ is the covariant derivative
(in adjoint representation). The first term 
in (\ref{2.4}) is the classical action (Pontryagin term) and the second one 
is the gauge--fixing term which removes the degeneracy of the classical 
action with respect to the topological shift symmetry $Q$ still leaving out 
of account the original gauge symmetry. Usually, $W_{\rm T}^{(\xi = 1)}$ is 
called equivariant part of the topological action being defined by the 
equivariant cohomology of TYM (prior introduction of the gauge ghost field).
Obviously, the Wess-Zumino gauge of supersymmetry immediatly leads to the
Feynman type gauge of TYM which is mediated by the vector charge 
$\bar{Q}_\mu$. Of course, other gauges may be used as well, e.g., those
introduced by Eqs.~(\ref{2.5}) below.
For notational simplicity we have set the gauge coupling 
equal to one, $g=1$. Let us also remark that, in principle,
the Pontryagin term could be multiplied by $\theta$ with some 
arbitrary parameter $\theta$ which, here, is set equal to one,
$\theta = 1$.

From the third relation in (\ref{2.2}) it follows that also the operator 
$\bar{Q}_\mu$ is equivariantly nilpotent, but now modulo the gauge 
transformation $\delta_G(\bar{\phi})$, and, for the same reason as before, 
that the field $\bar{\phi}$ must be $\bar{Q}_\mu$--invariant, 
$\bar{Q}_\mu \bar{\phi} = 0$. Let us stress that the second  
term in (\ref{2.4}) is different from zero if and only if $\bar{Q}_\mu$ does 
not anticommute with itself, showing the importance of the presence of 
the gauge transformation on the right--hand side of 
$\{ \bar{Q}_\mu, \bar{Q}_\nu \} = - 2 \delta_{\mu\nu} \delta_G(\bar{\phi})$. 
On the other hand, that gauge--fixing in (\ref{2.4}) is incomplete. 
In order to obtain the complete action
of TYM one still has to add a further gauge--fixing term removing 
the degeneracy of the classical action with respect to the remaining 
gauge symmetry (which will be done in Subsect.~2.3). However, that additional 
term turns out to be 
invariant under the vector supersymmetry $\bar{Q}_\mu$ if and only if 
$\{ \bar{Q}_\mu, \bar{Q}_\nu \} = 0$, i.e., if $\bar{Q}_\mu$ is 
strictly nilpotent. Hence, by choosing a Feynman type gauge the
vector charge
$\bar{Q}_\mu$ {can not be} really a symmetry operator of the complete 
gauge--fixed action of TYM. However, such a situation can be circumvented 
by choosing a Landau type gauge, i.e., by modifying the second term of the
action (\ref{2.4}) appropriately. 

In order to prepare the frame for a more general covariant gauge let us 
deform the topological superalgebra (\ref{2.2}) without changing its
topological character:
\begin{equation}
\label{2.5}
\{ Q, Q \} = - 2 \delta_G(\phi),
\qquad\!\!\!
\{ Q, \bar{Q}_\mu^{(\xi)} \} = - i P_\mu + \delta_G(A_\mu),
\qquad\!\!\!
\{ \bar{Q}_\mu^{(\xi)}, \bar{Q}_\nu^{(\xi)} \} = - 
2 \xi \delta_{\mu\nu} \delta_G(\bar{\phi}),
\end{equation}
thus making $\bar{Q}_\mu $ also $\xi$--dependent.
Here, $\xi$ is the gauge parameter interpolating between
Feynman ($\xi = 1$) and Landau ($\xi = 0$) type gauge. In turn, 
the topological action (\ref{2.4}) changes into
\begin{equation}
\label{2.6}
W_{\rm T}^{(\xi)} = \frac{1}{4 \xi} \int d^4x\, {\rm tr}\,
F_{\mu\nu} \tilde{F}^{\mu\nu} - \frac{1}{24 \xi} \epsilon^{\mu\nu\rho\sigma} 
\bar{Q}_\mu^{(\xi)} \bar{Q}_\nu^{(\xi)} \bar{Q}_\rho^{(\xi)}
\bar{Q}_\sigma^{(\xi)} \int d^4x\, {\rm tr}\, \phi^2, 
\end{equation}
where both terms are rescaled by $1/\xi$ in order to ensure 
that the action is well--defined also for $\xi = 0$ (see Eq.~(\ref{2.9}) 
below). From (\ref{2.6}) it follows that, by construction, 
$W_{\rm T}^{(\xi)}$ will be left invariant by the twisted operators 
$Q$ and $\bar{Q}_\mu^{(\xi)}$,
\begin{equation*}
Q W_{\rm T}^{(\xi)} = 0,
\qquad
\bar{Q}_\mu^{(\xi)} W_{\rm T}^{(\xi)} = 0.
\end{equation*}

The symmetry transformations corresponding to $Q$ and $\bar{Q}_\mu^{(\xi)}$ 
can be obtained by applying the 
previous twisting procedure to the transformation laws generated by 
$Q_A^{\!~~a}$ and $\bar{Q}_{\dot{A} a}$ (see Eqs.~(B.6) for the on--shell
transformations in the Feynman type gauge). 
Before showing their explicit form let us briefly introduce the twisted gauge 
multiplet $V_T$ of TYM by its relation to the 
gauge multiplet $V = \{ A_\mu, \lambda_A^{\!~~a}, \bar{\lambda}_{\dot{A} a}, 
\phi, \bar{\phi}, \chi_{ab} \}$ of $N = 2$ SYM. 

The anti--chiral spinor $\bar{\lambda}_{\dot{A} a}$, 
being the hermitean conjugate of $\lambda_A^{\!~~a}$, is related to a 
Grassmann--odd vector field $\psi_\mu$ through
\begin{equation*}
\psi_\mu = - \hbox{\large$\frac{1}{2}$}
 (\sigma_\mu)^{\dot{A} B} \bar{\lambda}_{\dot{A} B},
\end{equation*}
which is the topological ghost; $\phi$ is the ghost of the topological ghost 
and $\bar{\phi}$ is the corresponding antighost. The chiral spinor 
$\lambda_A^{\!~~a}$ is associated with both a Grassmann--odd antisymmetric 
self--dual tensor field $\chi_{\mu\nu}\equiv \chi_{\mu\nu}^+$ 
and a Grassmann--odd scalar field $\eta$ according to 
\begin{equation*}
\chi_{\mu\nu} = - \hbox{\large$\frac{1}{2}$}
i (\sigma_{\mu\nu})^{AB} \lambda_{AB},
\qquad
\eta = - \hbox{\large$\frac{1}{2}$} i \epsilon^{AB} \lambda_{AB},
\end{equation*}
where $\eta$ plays the role of an auxiliary field. Finally,
the symmetric auxiliary field $\chi_{ab}$ 
is related to a Grassmann--even antisymmetric 
self--dual tensor field $\lambda_{\mu\nu} \equiv \lambda_{\mu\nu}^+$ through
\begin{equation*}
\lambda_{\mu\nu} = \hbox{\large$\frac{1}{2}$}
(\sigma_{\mu\nu})^{AB} \chi_{AB},
\end{equation*}
which, again, ensures the closure of the twisted superalgebra (\ref{2.5}).  
Hence, the field content of TYM is given by the following twisted gauge 
multiplet $V_T = \{ A_\mu, \psi_\mu, \chi_{\mu\nu},\eta, \phi, \bar{\phi},  
\lambda_{\mu\nu}\}$ whose properties, together with those of 
the topological charges, are displayed in the 
following Table 1.
\bigskip
\begin{equation*}
\hbox{
\setlength{\extrarowheight}{3pt}
\begin{tabular}{|c|c|c|c|c|c|c|c||c|c|}
\hline
& $A_\mu$ & $\psi_\mu$ & $\chi_{\mu\nu}$ & $\eta$ & $\lambda_{\mu\nu}$ 
& $\phi$ & $\bar{\phi}$ & $Q$ & $\bar{Q}_\mu$
\\ 
\hline
ghost number & 0 & 1 & --1 & --1 & 0 & 2 & --2  & 1 & --1
\\ 
\hline
mass dimension & 1 & 3/2 & 3/2 & 3/2 & 2 & 1 & 1 & 1/2 & 1/2
\\ 
\hline
scale dimension & 1 & 1 & 2 & 2 & 2 & 0 & 2 & 0 & 1
\\ 
\hline
$Z$--, $\bar{Z}$--charge & 0 & 0 & 0 & 0 & 0 & 0 & 0 & 0 & 0
\\ 
\hline
Grassmann parity & even & odd & odd & odd & even &even & even & odd & odd
\\
\hline
\end{tabular}}
\end{equation*}

\bigskip
\bigskip

Let us now give the transformation law for the twisted gauge multiplet:\\
(i)
The topological shift symmetry $Q$ takes the form
\begin{align}
\label{2.7}
Q A_\mu &= \psi_\mu,
\qquad
Q \psi_\mu = D_\mu \phi,
\qquad
Q \phi = 0,
\nonumber
\\
Q \bar{\phi} &= \eta,
\qquad
Q \eta = [ \bar{\phi}, \phi ],
\qquad
Q \chi_{\mu\nu} = \lambda_{\mu\nu},
\qquad
Q \lambda_{\mu\nu} = [ \chi_{\mu\nu}, \phi ];
\end{align}
it is independent of the choice of $\xi$. Indeed, 
after rescaling by $\xi$ the fields $\bar{\phi}$, $\eta$ and 
$\chi_{\mu\nu}$, $\lambda_{\mu\nu}$ --- which belong to the non--minimal 
part (in the sense of BRST symmetry) of the multiplet $V_T$ --- 
these transformation rules obviously remain unchanged. \\
(ii)
The transformations rules for the vector supersymmetry 
$\bar{Q}_\mu^{(\xi)}$, are given by
\begin{align}
\label{2.8}
\bar{Q}_\mu^{(\xi)} A_\nu &= \xi ( \delta_{\mu\nu} \eta + \chi_{\mu\nu} ),
\qquad 
\bar{Q}_\mu^{(\xi)} \psi_\nu = F_{\mu\nu} - 
\xi ( \delta_{\mu\nu} [ \bar{\phi}, \phi ] + \lambda_{\mu\nu} ),
\qquad
\bar{Q}_\mu^{(\xi)} \phi = \psi_\mu, 
\nonumber
\\
\bar{Q}_\mu^{(\xi)} \bar{\phi} &= 0,
\qquad
\bar{Q}_\mu^{(\xi)} \eta = D_\mu \bar{\phi},
\qquad
\bar{Q}_\mu^{(\xi)} \chi_{\rho\sigma} = \delta_{\mu\rho} D_\sigma \bar{\phi} -
\delta_{\mu\sigma} D_\rho \bar{\phi} +
\epsilon_{\mu\nu\rho\sigma} D^\nu \bar{\phi},
\nonumber
\\
\bar{Q}_\mu^{(\xi)} \lambda_{\rho\sigma} &= D_\mu \chi_{\rho\sigma} +
\delta_{\mu\rho} ( [ \bar{\phi}, \psi_\sigma ] - D_\sigma \eta ) -
\delta_{\mu\sigma} ( [ \bar{\phi}, \psi_\rho ] - D_\rho \eta ) +
\epsilon_{\mu\nu\rho\sigma} ( [ \bar{\phi}, \psi^\nu ] - D^\nu \eta ).
\end{align} 
These transformations sensitively
depend on the choice of the gauge parameter $\xi$. Notice, that for 
$\xi = 0$ the operator $\bar{Q}_\mu^{(\xi)}$ leaves the gauge 
field $A_\mu$ inert and, therefore, does not change the classical action. 
Thus, the vector supersymmetry represents a non--trivial 
symmetry only with respect the gauge--fixing terms of TYM. 

Now, we are in a position to determine the second term in the topological
action (\ref{2.6}) explicitely. As a result one gets
(remind that $\lambda^{\mu\nu}$ and $\chi^{\mu\nu}$ are self-dual)
\begin{align}
\label{2.9}
W_{\rm T}^{(\xi)} = \int d^4x\, {\rm tr}\Bigr\{&
\lambda^{\mu\nu} F_{\mu\nu} - 2 \chi^{\mu\nu} D_\mu \psi_\nu + 
2 \eta D^\mu \psi_\mu + 2 \bar{\phi} \{ \psi^\mu, \psi_\mu \} + 
2 \bar{\phi} D^2 \phi
\\
& - \xi \Bigr(
\hbox{\large$\frac{1}{2}$} \lambda^{\mu\nu} \lambda_{\mu\nu} -
\hbox{\large$\frac{1}{2}$} \chi^{\mu\nu} [ \chi_{\mu\nu}, \phi ] +
2 [ \bar{\phi}, \phi ] [ \bar{\phi}, \phi ] -
2 \eta [ \eta, \phi ] \Bigr) \Bigr\}.
\nonumber
\end{align}
The relative factor in front of both terms of $W_{\rm T}^{(\xi)}$ 
in Eq.~(\ref{2.6}) (as well as (\ref{2.4})) has 
been choosen in such a way that the Pontryagin term is cancelled by the
contributions of the gauge--fixing term and that the remainder can 
be written as an exact $Q$--cocycle,
\begin{equation}
\label{2.10}
W_{\rm T}^{(\xi)} = Q \Psi_{\rm T}^{(\xi)}
\qquad
\hbox{with}
\qquad
\bar{Q}_\mu^{(\xi)} \Psi_{\rm T}^{(\xi)} = 0,
\end{equation}
where $\Psi_{\rm T}^{(\xi)}$ is the gauge fermion,
\begin{equation}
\label{2.11}
\Psi_{\rm T}^{(\xi)} = \int d^4x\, {\rm tr}\Bigr\{
\chi^{\mu\nu} F_{\mu\nu} + 2 \bar{\phi} D^\mu \psi_\mu - \xi \Bigr(
\hbox{\large$\frac{1}{2}$} \chi^{\mu\nu} \lambda_{\mu\nu} + 
2 \eta [ \bar{\phi}, \phi ] \Bigr) \Bigr\}. 
\end{equation}
This is that crucial property which allows to interpret TYM as a 
cohomological theory \cite{2}. Furthermore, let us stress that by 
imposing the vector supersymmetry $\bar{Q}_\mu^{(\xi)} W_{\rm T}^{(\xi)} = 0$, 
which implies $\bar{Q}_\mu^{(\xi)} \Psi_{\rm T}^{(\xi)} = 0$, the 
gauge fermion $\Psi_{\rm T}^{(\xi)}$ is determined uniquely.


\begin{flushleft}
{2.2. {\it Massive topological QCD: Equivariant part}}
\end{flushleft}
\bigskip
So far we have implicitely assumed that the complex field $\phi$ does not 
induce a central extension of the SUSY algebra (\ref{2.1}), or
equivalently, that its vacuum expectation value is zero \cite{24}. Now we 
remove that restriction and consider $N = 2$ SYM coupled to a massive
hypermultiplet which, after twisting, leads to topological QCD.

In this more general case we are faced with a 
$N = 2$ SUSY algebra with two central charges $Z$ and $\bar Z$,
\begin{align}
\label{2.13}
\{ Q_A^{\!~~a}, Q_B^{\!~~b} \} &= 
- 4 \epsilon^{ab} \epsilon_{AB} ( Z + \delta_G(\phi) ),
\nonumber
\\
\{ Q_A^{\!~~a}, \bar{Q}_{\dot{B} b} \} &=  
- 2 \delta^a_b (\sigma^\mu)_{A \dot{B}} ( P_\mu + i \delta_G(A_\mu) ),
\\
\{ \bar{Q}_{\dot{A} a}, \bar{Q}_{\dot{B} b} \} &= \phantom{-}
4 \epsilon_{ab} \epsilon_{\dot{A} \dot{B}} ( \bar{Z} + \delta_G(\bar{\phi}) ),
\nonumber
\end{align}
$\bar{Z}$ being the complex conjugate of $Z$. Thereby, $Z$ and $\bar{Z}$, 
together with $P_\mu$, satisfy the condition $4 Z \bar{Z} = P^\mu P_\mu$. 
The eigenvalues of $Z$ and $\bar{Z}$ are $\pm m$
and $\pm\bar{m}$, respectively, where the positive (negative) sign relates
to the hypermultiplet $Y \,(\bar{Y}$). 
Here, we have introduced two central charges in order to ensure that, 
lateron, the $R$--symmetry remains unbroken if we {\em formally} assign 
to $Z$ and $\bar{Z}$ the $R$--charges $2$ and $-2$, respectively. 
(By coupling the gauge multiplet to the standard massive hypermultiplet 
\cite{25}, i.e., for only one (real) central charge $Z = \bar{Z}$, 
the $R$--symmetry is broken into $Z_2$; cf.,~Ref.~\cite{4}.)  

In order to construct the topological action with central charges $Z$ and
$\bar{Z}$ we may consider in a $D=6$ dimensional space--time \cite{22}
a $N = 1$ SYM whose gauge multiplet $\{A_M, \lambda_a \}$ is coupled to 
a (massless) hypermultiplet $\{\psi, \zeta_a\}$ in some representation
$\cal R$ of the gauge group $G$ (for a detailed presentation, see,
Appendix B).  Then, one carries out a non--trivial dimensional reduction 
onto a torus with respect to the extra spatial dimensions $x^5$ and $x^6$ 
\cite{26}, thereby, introducing two (real) masses, 
$m^5 = m + \bar{m}$ and $m^6 = i ( \bar{m} - m )$, which are the inverse
periods of the fields of the hypermultiplet with respect to $x^5$ and $x^6$.
The extra components $P_5$ and $P_6$ of the generator of space--time 
translation are related to the central charges according to 
$P_5 = Z + \bar{Z} $ and $P_6 = i ( \bar{Z} - Z)$, respectively. In the 
same way, the extra components $A_5$ and $A_6$ of the gauge field,
which are assumed to be independent on $x^5$ and $x^6$, are 
identified with the complex scalar field according to 
$A_5 = - i (\bar{\phi} + \phi)$ and $A_6 = \bar{\phi} - \phi$.  

After compactification, assuming the only the first non--zero modes are 
excited, the hypermultiplet $Y$ (and its hermitean conjugate $\bar Y$) 
becomes massive. The resulting hypermultiplet $Y$ consists of two Weyl 
spinors being the components of the corresponding (Dirac) spinor, 
$q = (\alpha^A, \bar{\beta}_{\dot{A}})^T$, a $Sp(2)$--doublet of complex 
scalar fields $\zeta^a$ and, in order to close the superalgebra (\ref{2.13})
also off--shell, a $Sp(2)$--doublet of complex 
auxiliary fields $\chi^a$.  After the twisting procedure the scalar 
fields $\zeta^a$ become the components of a bispinor field $\zeta^A$. 
The appearance of bispinors after twisting is a new feature of TQCD. 
Similarly, the auxiliary field $\chi^a$ should be replaced by a bispinor 
field $\chi^A$. But, a further and somehow surprising feature of the 
twisting procedure is that, in the presence of central charges,  
twisting of the superalgebra (\ref{2.13}) does not lead to 
a topological superalgebra which closes off--shell \cite{27}
-- despite having started with a closed algebra! 

However, one can proceed like in the case of twisting $N = 2$ conformal 
supergravity \cite{28}. Namely, replacing $\chi^a$ instead by $\chi^A$
through another bispinor $\bar{\chi}_{\dot{A}}$ the resulting 
topological superalgebra closes off--shell \cite{16}. 
Thus, the multiplets from which TQCD is constructed consist of the 
(massless) twisted gauge multiplet $V_T = \{ 
A_\mu, \psi_\mu, \chi_{\mu\nu}, \eta, \phi, \bar{\phi}, 
\lambda_{\mu\nu} \}$ in the adjoint representation of $G$ and two  
twisted massive hypermultiplets 
$Y_T = \{\alpha^A, \bar{\beta}_{\dot{A}}, \zeta^A, \bar{\chi}_{\dot{A}}\}$
and $\bar{Y}_T = \{
\beta_A, \bar{\alpha}^{\dot{A}}, \bar{\zeta}_A, \chi^{\dot{A}} \}$
in the representation $\cal R$. 
Their properties are displayed in the following Table 2. 
\bigskip
\begin{equation*}
\hbox{
\setlength{\extrarowheight}{3pt}
\begin{tabular}{|c|c|c|c|c|c|c|c|c|}
\hline
& $\alpha^A$ & $\beta_A$ & 
$\bar{\alpha}^{\dot{A}}$ & $\bar{\beta}_{\dot{A}}$ 
& $\zeta^A$ & $\bar{\zeta}_A$ 
& $\chi^{\dot{A}}$ & $\bar{\chi}_{\dot{A}}$ 
\\ 
\hline
ghost number & 1 & 1 & --1 & --1 & 0 & 0 & 0 & 0 
\\ 
\hline
mass dimension & 3/2 & 3/2 & 3/2 & 3/2 & 1 & 1 & 2 & 2 
\\ 
\hline
scale dimension & 1 & 1 & 2 & 2 & 1 & 1 & 2 & 2 
\\ 
\hline
$Z$--, $\bar{Z}$--charge & 1 & --1 & --1 & 1 & 
                           1 & --1 & 1 & --1 
\\ 
\hline
Grassmann parity & even & even & even & even & odd & odd & odd & odd 
\\
\hline
\end{tabular}}
\end{equation*}
\medskip

The twisted action for the matter fields, including the terms for the
new auxiliarly fields $\chi^{\dot{A}}$ and $\bar{\chi}_{\dot{A}}$, is
\begin{align}
\label{2.14}
W_{\rm M}^{(\xi)} = \int d^4x\, \Bigr\{&
i \bar{\alpha}^{\dot{A}} (\sigma^\mu)_{\dot{A} B} 
\overset{\rightarrow}{D}_\mu \alpha^B +
i \beta_A \overset{\leftarrow}{D}_\mu (\sigma^\mu)^{A \dot{B}} 
\bar{\beta}_{\dot{B}} - 2 \bar{m} \beta_A \alpha^A 
\nonumber
\\
& + i \bar{\alpha}^{\dot{A}} (\sigma^\mu)_{\dot{A} B} \psi_\mu \zeta^B - 
i \bar{\zeta}_A (\sigma^\mu)^{A \dot{B}} \psi_\mu \bar{\beta}_{\dot{B}} - 
2 \bar{\alpha}^{\dot{A}} ( \phi + m ) \bar{\beta}_{\dot{A}} 
\nonumber
\\
\phantom{\frac{1}{2}}
& - i \chi^{\dot{A}} (\sigma^\mu)_{\dot{A} B} 
\overset{\rightarrow}{D}_\mu \zeta^B +
i \bar{\zeta}_A \overset{\leftarrow}{D}_\mu (\sigma^\mu)^{A \dot{B}} 
\bar{\chi}_{\dot{B}} - 2 \chi^{\dot{A}} \bar{\chi}_{\dot{A}} + 
2 \bar{m} \bar{\zeta}_A ( \phi + m ) \zeta^A 
\\
& - \xi \Bigr( 
\hbox{\large$\frac{1}{4}$} \beta_A (\sigma^{\mu\nu})^{AB} 
\chi_{\mu\nu} \zeta_B + 
\hbox{\large$\frac{1}{4}$} \bar{\zeta}_A (\sigma^{\mu\nu})^{AB} 
\lambda_{\mu\nu} \zeta_B -
\hbox{\large$\frac{1}{4}$} \bar{\zeta}_A (\sigma^{\mu\nu})^{AB} 
\chi_{\mu\nu} \alpha_B
\nonumber
\\
& + \bar{\zeta}_A \eta \alpha^A - 
\bar{\zeta}_A \bar{\phi} ( \phi + m ) \zeta^A + 
2 \beta_A \bar{\phi} \alpha^A + 
\beta_A \eta \zeta^A - 
\bar{\zeta}_A ( \phi + m ) \bar{\phi} \zeta^A \Bigr) \Bigr\},
\nonumber
\end{align}
where the covariant derivatives in the $\cal R$--representation are
given by
\begin{equation*}
\overset{\rightarrow}{D}_\mu = \overset{\rightarrow}{\partial}_\mu + A_\mu,
\qquad
\overset{\leftarrow}{D}_\mu = \overset{\leftarrow}{\partial}_\mu - A_\mu.
\end{equation*}
In the Feynman type gauge ($\xi = 1$) this part of the action is symmetric in
$m$ and $\bar m$. Furthermore, let us recall that in the twisted theory 
the $R$--symmetry is identified with 
the ghost number conservation. Hence, by inspection of (\ref{2.14}), we 
observe that the mass terms preserve the ghost number if we formally assign 
to $m$ and $\bar{m}$ the ghost number $2$ and $-2$, respectively. 

Let us now consider the transformation law of the twisted  
hypermultiplets. \\
(i)
The topological shift symmetry $Q$ takes the form
\begin{alignat}{4}
\label{2.15}
Q \zeta^A &= \alpha^A,
&\qquad
Q \alpha^A & = - ( \phi + m ) \zeta^A,
&\qquad
Q \bar{\beta}_{\dot{A}} &= \bar{\chi}_{\dot{A}},
&\qquad
Q \bar{\chi}_{\dot{A}} &= - ( \phi + m ) \bar{\beta}_{\dot{A}},
\nonumber
\\
Q \bar{\zeta}_A &= \beta_A,
&\qquad
Q \beta_A & = \bar{\zeta}_A ( \phi + m ),
&\qquad
Q \bar{\alpha}^{\dot{A}} &= \chi^{\dot{A}},
&\qquad
Q \chi^{\dot{A}} &= \bar{\alpha}^{\dot{A}} ( \phi + m );
\end{alignat}
again, it is independent of the choice of the gauge parameter $\xi$. 
But now, the anticommutator of the operator $Q$ includes, besides the 
gauge transformation $- 2 \delta_G(\phi)$, for $m \neq 0$ also the 
central charge transformation $Z$. Together with those of $\bar{Z}$ 
they are given by    
\begin{align}
\label{2.16}
Z V_T = 0,
\qquad
Z Y_T& = m Y_T,
\qquad
Z \bar{Y}_T = - m \bar{Y}_T,
\nonumber
\\
\bar{Z} V_T = 0,
\qquad
\bar{Z} Y_T& = \bar{m} Y_T,
\qquad
\bar{Z} \bar{Y}_T = - \bar{m} \bar{Y}_T.
\end{align}
(ii)
The transformation rules for the vector supersymmetry 
$\bar{Q}_\mu^{(\xi)}$ are given by 
\begin{align}
\label{2.17}
\bar{Q}_\mu^{(\xi)} \zeta^A &= 
i (\sigma_\mu)^{A \dot{B}} \bar{\beta}_{\dot{B}},
\nonumber
\\
\bar{Q}_\mu^{(\xi)} \bar{\beta}_{\dot{A}} &= i (\sigma_\mu)_{\dot{A} B} 
( \xi \bar{\phi} + \bar{m} ) \zeta^B,
\nonumber
\\
\bar{Q}_\mu^{(\xi)} \alpha^A &= 
- i (\sigma_\mu)^{A \dot{B}} \bar{\chi}_{\dot{B}} + 
\overset{\rightarrow}{D}_\mu \zeta^A,
\nonumber
\\
\bar{Q}_\mu^{(\xi)} \bar{\chi}_{\dot{A}} &= 
- i (\sigma_\mu)_{\dot{A} B} ( \xi \bar{\phi} + \bar{m} ) \alpha^B - 
\xi i (\sigma_\mu)_{\dot{A} B} \eta \zeta^B + 
\overset{\rightarrow}{D}_\mu \bar{\beta}_{\dot{A}},
\nonumber
\\
\bar{Q}_\mu^{(\xi)} \bar{\zeta}_A &= 
- i (\sigma_\mu)_{A \dot{B}} \bar{\alpha}^{\dot{B}},
\nonumber
\\
\bar{Q}_\mu^{(\xi)} \bar{\alpha}^{\dot{A}} &= i (\sigma_\mu)^{\dot{A} B}
\bar{\zeta}_B ( \xi \bar{\phi} + \bar{m} ),
\nonumber
\\
\bar{Q}_\mu^{(\xi)} \beta_A &= 
i (\sigma_\mu)_{A \dot{B}} \chi^{\dot{B}} + 
\bar{\zeta}_A \overset{\leftarrow}{D}_\mu,
\nonumber
\\
\bar{Q}_\mu^{(\xi)} \chi^{\dot{A}} &= 
- i (\sigma_\mu)^{\dot{A} B} \beta_B ( \xi \bar{\phi} + \bar{m} ) -
\xi i (\sigma_\mu)^{\dot{A} B} \bar{\zeta}_B \eta + 
\bar{\alpha}^{\dot{A}} \overset{\leftarrow}{D}_\mu.
\end{align}
As before, the anticommutator $\{\bar{Q}_\mu^{(\xi)}, \bar{Q}_\nu^{(\xi)}\}$
does no longer coincide with the $\xi$--dependent gauge transformation 
$- 2 \xi \delta_{\mu\nu} \delta_G(\bar{\phi})$ alone but in addition 
it includes the (complex conjugate) central charge transformation $\bar{Z}$.

By a rather lengthy calculation it can be proven that the transformations 
(\ref{2.15}) -- (\ref{2.17}) together with (\ref{2.7}) and (\ref{2.8}) 
satisfy the following topological superalgebra:
\begin{align}
\label{2.18}
\{ Q, Q \} &= - 2 ( Z + \delta_G(\phi) ),
\nonumber\\
\{ Q, \bar{Q}_\mu^{(\xi)} \} &= -iP_\mu + \delta_G(A_\mu),
\\
\{ \bar{Q}_\mu^{(\xi)}, \bar{Q}_\nu^{(\xi)} \} &= - 2 \delta_{\mu\nu} 
( \bar{Z} + \xi \delta_G(\bar{\phi}) ).
\nonumber
\end{align}

By making use of the $Q$--transformations (\ref{2.15}) it is easy to show 
that the twisted action $W_{\rm M}^{(\xi)}$ can be written as an exact 
$Q$--cocycle, 
\begin{equation}
\label{2.19}
W_{\rm M}^{(\xi)} = Q \Psi_{\rm M}^{(\xi)}
\qquad
\hbox{with}
\qquad
\bar{Q}_\mu^{(\xi)} \Psi_{\rm M}^{(\xi)} = 0,
\qquad
Z \Psi_{\rm M}^{(\xi)} = 0,
\qquad
\bar{Z} \Psi_{\rm M}^{(\xi)} = 0,
\end{equation}
where the $\bar{m}$--dependent gauge fermion $\Psi_{\rm M}^{(\xi)}$ is given by
\begin{align}
\label{2.20}
\Psi_{\rm M}^{(\xi)} = - \int d^4x\, \Bigr\{&
i \bar{\alpha}^{\dot{A}} (\sigma^\mu)_{\dot{A} B} 
\overset{\rightarrow}{D}_\mu \zeta^B -
i \bar{\zeta}_A \overset{\leftarrow}{D}_\mu (\sigma^\mu)^{A \dot{B}} 
\bar{\beta}_{\dot{B}} +
\bar{m} ( \bar{\zeta}_A \alpha^A - \beta_A \zeta^A )
\nonumber
\\
& + \bar{\alpha}^{\dot{A}} \bar{\chi}_{\dot{A}} +
\chi^{\dot{A}} \bar{\beta}_{\dot{A}} + \xi \Bigr(
\bar{\zeta}_A \bar{\phi} \alpha^A - \beta_A \bar{\phi} \zeta^A +
\frac{1}{4} \bar{\zeta}_A 
(\sigma^{\mu\nu})^{AB} \chi_{\mu\nu} \zeta_B \Bigr) \Bigr\}.
\end{align}
Let us stress that the gauge fermion $\Psi_{\rm M}^{(\xi)}$ is  
completely specified by the topological shift symmetry $Q$ 
and the vector supersymmetry ${\bar Q}_\mu^{(\xi)}$. Furthermore,
identifying $Z=\bar Z$ and, therefore, $m = \bar m$, we obtain for the
matter action, Eq.~(\ref{2.14}), the result of \cite{15}.

Hence, the twisted $N = 2$ SYM coupled to the massive hypermultiplets 
$Y_T$ and $\bar{Y}_T$ is described by the action
\begin{equation}
\label{2.21}
W^{(\xi)} = W_{\rm T}^{(\xi)} + W_{\rm M}^{(\xi)} = 
Q ( \Psi_{\rm T}^{(\xi)} + \Psi_{\rm M}^{(\xi)} ).
\end{equation} 
This defines completely the equivariant part of TQCD.
It is invariant under the topological superalgebra (\ref{2.18}) and
possesses the crucial property of being an exact $Q$--cocycle.


\bigskip
\begin{flushleft}
{2.3. {\it Massive topological QCD: Complete action in Landau gauge}}
\end{flushleft}
\bigskip
Next, we are faced with the problem to construct the complete action 
of TQCD. This is achieved by adding to the equivariant part 
a supplementary gauge--fixing term which removes the residual gauge 
degeneracy of the classical action thereby preserving the invariance under
the topological superalgebra (\ref{2.18}). The last requirement
turns out to be very restrictive and can be fulfilled only when in 
(\ref{2.21}) the Landau type gauge $\xi = 0$ has been chosen.

To begin with, let us introduce the following prolonged BRST operator
\begin{equation}
\label{2.22}
s_Q = s + Q,
\end{equation}
which, according to the first of Eqs.~(\ref{2.18}), 
is required to be nilpotent modulo the central charge $Z$, 
i.e., $\{ s_Q, s_Q \} = - 2 Z$. In order to identify $s$ with the (usual)
nilpotent BRST operator, let us introduce the gauge ghost $C$ by assuming, 
as usual, that the topological ghost $\phi$ is the exact 
$Q$--cocycle, $\phi = Q C$. Then, applying 
$\{ Q, Q \} = - 2 ( Z + \delta_G(\phi) )$ on $C$ it 
follows that $s$ is defined by $s = \delta_G(C)$ when acting on the gauge 
multiplet $V$ and $s C = C^2$ when acting on the gauge ghost. The 
transformation law for the antighost $\bar{C}$ and the auxiliary field 
$B$ will be defined by $s \bar{C} = B + \{ C, \bar{C} \}$ and 
$s B = [ \bar{C}, \phi ] + [ C, B ]$. Notice, that $C$, $\bar{C}$ and $B$ 
take their values in $Lie(G)$ and have vanishing
central charges; their properties are summarized in the following Table 3.
\bigskip
\begin{equation*}
\hbox{
\setlength{\extrarowheight}{3pt}
\begin{tabular}{|c|c|c|c|}
\hline
& $C$ & $\bar{C}$ & $B$  
\\ 
\hline
ghost number & 1 & --1 & 0  
\\ 
\hline
mass dimension & 1/2 & 3/2 & 2  
\\ 
\hline
scale dimension & 0 & 2 & 2  
\\ 
\hline
$Z$--, $\bar{Z}$--charge & 0 & 0 & 0  
\\ 
\hline
Grassmann parity & odd & odd & even
\\
\hline
\end{tabular}}
\end{equation*}
\bigskip

Together with the action of the topological shift symmetry $Q$, 
Eqs.~(\ref{2.7}), we get the BRST transformation:
\begin{alignat}{2}
\label{2.25}
s_Q A_\mu &= \psi_\mu - D_\mu C,
&\qquad
s_Q \psi_\mu &= D_\mu \phi + \{ C, \psi_\mu \},
\nonumber
\\
s_Q C &= \phi + C^2,
&\qquad
s_Q \phi &= [ C, \phi ],
\nonumber
\\
s_Q \bar{\phi} &= \eta + [ C, \bar{\phi} ],
&\qquad
s_Q \eta &= [ \bar{\phi}, \phi ] + \{ C, \eta \},
\nonumber
\\
s_Q \chi_{\mu\nu} &= \lambda_{\mu\nu} + \{ C, \chi_{\mu\nu} \},
&\qquad
s_Q \lambda_{\mu\nu} &= [ \chi_{\mu\nu}, \phi ] + [ C, \lambda_{\mu\nu} ],
\nonumber
\\
s_Q \bar{C} &= B + \{ C, \bar{C} \},
&\qquad
s_Q B &= [ \bar{C}, \phi ] + [ C, B ],
\end{alignat}

Now, we are in a position to define the complete action of TYM by
adding to the topological action (\ref{2.10}) and (\ref{2.11}) the 
remaining gauge fixing part $W_{\rm YM}$,
\begin{equation}
\label{2.23}
W_{\rm TYM}^{(\xi)} = W_{\rm T}^{(\xi)} + W_{\rm YM} 
\qquad 
{\rm with}
\qquad
W_{\rm T}^{(\xi)} = s_Q \Psi_{\rm T}^{(\xi)},
\qquad
W_{\rm YM}=s_Q \Psi_{\rm YM},
\end{equation}
where $\Psi_{\rm YM}$ is the YM gauge fermion in the Landau gauge,
\begin{equation}
\label{2.24}
\Psi_{\rm YM} = 2 \int d^4x\, {\rm tr} \Bigr\{ 
\bar{C} \partial^\mu A_\mu \Bigr\}. 
\end{equation}
(In principle, also here we could have choosen a gauge fermion 
$\Psi_{\rm YM}^{(\xi')}$ in an intermediate gauge with 
another parameter $\xi'$ which, for the Landau gauge, is set equal
to zero.)
Notice, that $W_{\rm T}^{(\xi)}$ can be rewritten in the form 
$W_{\rm T}^{(\xi)} = s_Q \Psi_{\rm T}^{(\xi)}$, 
since $\Psi_{\rm T}^{(\xi)}$ is gauge invariant, i.e., 
$s \Psi_{\rm T}^{(\xi)} = 0$. Hence, by construction, the action ({\ref{2.23}) 
is invariant under the modified BRST transformations, Eqs.~({\ref{2.25}).
For the irreducible part of TYM we obtain
\begin{equation}
\label{2.26}
W_{\rm YM} = s_Q \Psi_{\rm YM} = 2 \int d^4x\, {\rm tr} \Bigr\{
B \partial^\mu A_\mu - C \partial^\mu D_\mu \bar{C} -
\bar{C} \partial^\mu \psi_\mu \Bigr\}.
\end{equation}

Let us now study the invariance properties of $W_{\rm YM}$ under the 
$\bar{Q}_\mu^{(\xi)}$--transformations. 
By applying $\{ Q, \bar{Q}_\mu^{(\xi)} \} = - i P_\mu + \delta_G(A_\mu)$ 
on $C$ and using $Q C = \phi$ we get $\bar{Q}_\mu^{(\xi)} C = - A_\mu$. 
This allows to cast the previous relation into the form 
$\{ s_Q, \bar{Q}_\mu^{(\xi)} \} = - i P_\mu$. Then, by making use of this 
new relation, from the requirements $s_Q \bar{C} = B + \{ C, \bar{C} \}$ and 
$s_Q B = [ \bar{C}, \phi ] + [ C, B ]$ we get 
$\bar{Q}_\mu^{(\xi)} \bar{C} = 0$ and 
$\bar{Q}_\mu^{(\xi)} B = D_\mu \bar{C}$.
Now, recalling that $\bar{Q}_\mu^{(\xi)}$ leaves $A_\mu$ inert iff 
$\xi = 0$, see Eqs.~(\ref{2.8}), by applying 
$\{ \bar{Q}_\mu^{(\xi)}, \bar{Q}_\nu^{(\xi)} \} = 
- 2 \delta_{\mu\nu} ( \bar{Z} + \xi \delta_G(\bar{\phi}) )$
on $C$ and using $\bar{Q}_\mu^{(\xi)} C = - A_\mu$ we conclude that
the irreducible part $W_{\rm YM}$ can not be invariant under the vector 
supersymmetry $\bar{Q}_\mu^{(\xi)}$ as long as $\xi \neq 0$. 

For that reason, from now on we shall choose the Landau type gauge,
$\xi = 0$, for the topological action. 
In that gauge the transformations of the complete gauge multiplet under
the vector supersymmetry $\bar{Q}_\mu^{(0)}$ read (cf.,~Eqs.~(\ref{2.8})):
\begin{align}
\label{2.27}
\bar{Q}_\mu^{(0)} A_\nu &= 0,
\qquad 
\bar{Q}_\mu^{(0)} \psi_\nu = F_{\mu\nu},
\nonumber
\\
\bar{Q}_\mu^{(0)} C &= - A_\mu,
\qquad
\bar{Q}_\mu^{(0)} \phi = \psi_\mu, 
\qquad
\bar{Q}_\mu^{(0)} B = D_\mu \bar{C},
\qquad
\bar{Q}_\mu^{(0)} \bar{C} = 0,
\nonumber
\\
\bar{Q}_\mu^{(0)} \bar{\phi} &= 0,
\qquad
\bar{Q}_\mu^{(0)} \eta = D_\mu \bar{\phi},
\qquad
\bar{Q}_\mu^{(0)} \chi_{\rho\sigma} = \delta_{\mu\rho} D_\sigma \bar{\phi} -
\delta_{\mu\sigma} D_\rho \bar{\phi} +
\epsilon_{\mu\nu\rho\sigma} D^\nu \bar{\phi},
\nonumber
\\
\bar{Q}_\mu^{(0)} \lambda_{\rho\sigma} &= D_\mu \chi_{\rho\sigma} +
\delta_{\mu\rho} ( [ \bar{\phi}, \psi_\sigma ] - D_\sigma \eta ) -
\delta_{\mu\sigma} ( [ \bar{\phi}, \psi_\rho ] - D_\rho \eta ) +
\epsilon_{\mu\nu\rho\sigma} ( [ \bar{\phi}, \psi^\nu ] - D^\nu \eta ),
\end{align} 
and for the massive hypermultiplets we obtain (cf.,~Eqs.~(\ref{2.17}))
\begin{alignat}{2}
\label{2.35}
\bar{Q}_\mu^{(0)} \zeta^A &= 
i (\sigma_\mu)^{A \dot{B}} \bar{\beta}_{\dot{B}},
&\qquad
\bar{Q}_\mu^{(0)} \bar{\beta}_{\dot{A}} &=  
i \bar{m} (\sigma_\mu)_{\dot{A} B} \zeta^B,
\nonumber
\\
\bar{Q}_\mu^{(0)} \alpha^A &= 
- i (\sigma_\mu)^{A \dot{B}} \bar{\chi}_{\dot{B}} + 
\overset{\rightarrow}{D}_\mu \zeta^A,
&\qquad
\bar{Q}_\mu^{(0)} \bar{\chi}_{\dot{A}} &= 
- i \bar{m} (\sigma_\mu)_{\dot{A} B} \alpha^B + 
\overset{\rightarrow}{D}_\mu \bar{\beta}_{\dot{A}},
\nonumber
\\
\bar{Q}_\mu^{(0)} \bar{\zeta}_A &= 
- i (\sigma_\mu)_{A \dot{B}} \bar{\alpha}^{\dot{B}},
&\qquad
\bar{Q}_\mu^{(0)} \bar{\alpha}^{\dot{A}} &= 
i \bar{m} (\sigma_\mu)^{\dot{A} B} \bar{\zeta}_B,
\nonumber
\\
\bar{Q}_\mu^{(0)} \beta_A &= 
i (\sigma_\mu)_{A \dot{B}} \chi^{\dot{B}} +
\bar{\zeta}_A \overset{\leftarrow}{D}_\mu,
&\qquad
\bar{Q}_\mu^{(0)} \chi^{\dot{A}} &= 
- i \bar{m} (\sigma_\mu)^{\dot{A} B} \beta_B + 
\bar{\alpha}^{\dot{A}}\overset{\leftarrow}{D}_\mu,
\end{alignat}
The corresponding BRST transformations of the hypermultiplet read 
(cf.,~Eqs.~(\ref{2.15}))
\begin{alignat}{2}
\label{2.34}
s_Q \zeta^A &= \alpha^A + C \zeta^A,
&\qquad
s_Q \alpha^A &= - ( \phi + m ) \zeta^A + C \alpha^A,
\nonumber
\\
s_Q \bar{\beta}_{\dot{A}} &= \bar{\chi}_{\dot{A}} + C \bar{\beta}_{\dot{A}},
&\qquad
s_Q \bar{\chi}_{\dot{A}} &= - ( \phi + m ) \bar{\beta}_{\dot{A}} +
C \bar{\chi}_{\dot{A}},
\nonumber
\\
s_Q \bar{\zeta}_A &= \beta_A - \bar{\zeta}_A C,
&\qquad
s_Q \beta_A &= 
\bar{\zeta}_A ( \phi + m ) + \beta_A C,
\nonumber
\\
s_Q \bar{\alpha}^{\dot{A}} &= \chi^{\dot{A}} + \bar{\alpha}^{\dot{A}} C,
&\qquad
s_Q \chi^{\dot{A}} &= 
\bar{\alpha}^{\dot{A}} ( \phi + m ) - \chi^{\dot{A}} C.
\end{alignat}

By a tedious but straightforward calculation it can be proven that the
transformations (\ref{2.16}), (\ref{2.25}) and (\ref{2.27}) together with
(\ref{2.35}) and (\ref{2.34}) satisfy the following superalgebra:
\begin{equation}
\label{2.36}
\{ s_Q, s_Q \} = - 2 Z,
\qquad
\{ s_Q, \bar{Q}_\mu^{(0)} \} = - i P_\mu,
\qquad
\{ \bar{Q}_\mu^{(0)}, \bar{Q}_\nu^{(0)} \} = - 2 \delta_{\mu\nu} \bar{Z},
\end{equation}
which is obtained from (\ref{2.18}) after substituting $Q$ by $s_Q$ 
and putting $\xi$ equal to zero.

From (\ref{2.9}) for the topological action in the Landau type gauge we obtain
\begin{equation}
\label{2.31}
W_{\rm T}^{(0)} = \int d^4x\, {\rm tr}\Bigr\{
\lambda^{\mu\nu} F_{\mu\nu} - 2 \chi^{\mu\nu} D_\mu \psi_\nu + 
2 \eta D^\mu \psi_\mu + 2 \bar{\phi} D^2 \phi + 
2 \bar{\phi} \{ \psi^\mu, \psi_\mu \} \Bigr\},
\end{equation}
and from (\ref{2.14}) for the matter action we immediately get  
\begin{align}
\label{2.37}
W_{\rm M}^{(0)} = \int d^4x\, \Bigr\{&
i \bar{\alpha}^{\dot{A}} (\sigma^\mu)_{\dot{A} B} 
\overset{\rightarrow}{D}_\mu \alpha^B +
i \beta_A \overset{\leftarrow}{D}_\mu (\sigma^\mu)^{A \dot{B}} 
\bar{\beta}_{\dot{B}} - 2 \bar{m} \beta_A \alpha^A 
\\
& + i \bar{\alpha}^{\dot{A}} (\sigma^\mu)_{\dot{A} B} \psi_\mu \zeta^B - 
i \bar{\zeta}_A (\sigma^\mu)^{A \dot{B}} \psi_\mu \bar{\beta}_{\dot{B}} + 
2 \bar{m} \bar{\zeta}_A ( \phi + m ) \zeta^A  
\phantom{\frac{1}{2}}
\nonumber
\\
& - i \chi^{\dot{A}} (\sigma^\mu)_{\dot{A} B} 
\overset{\rightarrow}{D}_\mu \zeta^B +
i \bar{\zeta}_A \overset{\leftarrow}{D}_\mu (\sigma^\mu)^{A \dot{B}}
\bar{\chi}_{\dot{B}} - 2 \chi^{\dot{A}} \bar{\chi}_{\dot{A}} -
2 \bar{\alpha}^{\dot{A}} ( \phi + m ) \bar{\beta}_{\dot{A}} \Bigr\}.
\nonumber
\end{align}
Here, $W_{\rm M}^{(0)} $ can be rewritten in the form 
$W_{\rm M}^{(0)} = s_Q \Psi_{\rm M}^{(0)}$, since $\Psi_{\rm M}^{(0)}$ 
is gauge invariant, i.e., $s \Psi_{\rm M}^{(0)}  = 0$. 

After all, putting anything together, we arrive at the complete action of 
TQCD (in Landau gauge) we are looking for,
\begin{align}
\label{2.33}
W_{\rm TQCD}^{(0)}  = W_{\rm TYM}^{(0)}  + W_{\rm M}^{(0)} = 
s_Q (\Psi_{\rm T}^{(0)} + \Psi_{\rm YM}^{(0)} ) + s_Q \Psi_{\rm M}^{(0)},
\end{align}
with
\begin{align}
\label{2.xx}
\Psi_{\rm T}^{(0)} = \int d^4x\, {\rm tr}\Bigr\{&
\chi^{\mu\nu} F_{\mu\nu} + 2 \bar{\phi} D^\mu \psi_\mu \Bigr\}, 
\\
\label{2.yy}
\Psi_{\rm M}^{(0)} = - \int d^4x\, \Bigr\{&
i \bar{\alpha}^{\dot{A}} (\sigma^\mu)_{\dot{A} B} 
\overset{\rightarrow}{D}_\mu \zeta^B -
i \bar{\zeta}_A \overset{\leftarrow}{D}_\mu (\sigma^\mu)^{A \dot{B}} 
\bar{\beta}_{\dot{B}} 
\nonumber
\\
& + \bar{m} ( \bar{\zeta}_A \alpha^A - \beta_A \zeta^A ) + 
\bar{\alpha}^{\dot{A}} \bar{\chi}_{\dot{A}} +
\chi^{\dot{A}} \bar{\beta}_{\dot{A}}  \Bigr\}.
\end{align}
and $\Psi_{\rm YM}$ given by Eq.~(\ref{2.24}).
It is invariant under the topological superalgebra (\ref{2.36}) and
possesses the crucial property of being an exact $s_Q$--cocycle.


\bigskip
\begin{flushleft}
{\large{\bf 3. Non--covariant Landau type gauge}}
\end{flushleft}
\bigskip
In Ref. \cite{17} it has been shown that if instead of the gauge covariant 
choice, $D_\mu \psi^\mu = 0$, cf.,~Eq.~(\ref{2.xx}), the
non--covariant Landau type gauge, $\partial_\mu \psi^\mu = 0$, 
is imposed then the action of TYM obeys a larger set of 
global constraints than in the former case. 
In particular, it turns out that the associated vector supersymmetry, 
in the following denoted
simply by $\bar{Q}_\mu$, is {\em linearly} realized. Therefore, it
can be employed as an additional stability constraint in order to improve 
the ultraviolet behaviour of TYM, restraining the number of independent 
invariant counterterms. Motivated by that result, it will be shown that the 
same gauge can be imposed also in the case of TQCD and that, except 
for the so--called ghost for the ghost equation \cite{17}, there is an 
additional set of stability constraints which improve the finiteness 
properties displayed by this model.

To begin with, let us first express the BRST transformations 
(\ref{2.25}) by redefining the fields
$\eta$, $\lambda^{\mu\nu}$ and $B$, being only auxiliary ones, according to  
\begin{eqnarray*}
\eta \rightarrow \eta - [ C, \bar{\phi} ],
\qquad
\lambda^{\mu\nu} \rightarrow \lambda^{\mu\nu} - \{ C, \chi^{\mu\nu} \},
\qquad 
B \rightarrow B - \{ C, \bar{C} \},
\end{eqnarray*}
in such a way that all the fields belonging to the non--minimal
sector occure as trivial BRST--doublets, 
\begin{align}
\label{3.1}
s_Q A_\mu &= \psi_\mu - D_\mu C,
\quad
s_Q \psi_\mu = D_\mu \phi + \{ C, \psi_\mu \},
\qquad
s_Q C = \phi + C^2,
\quad
s_Q \phi = [ C, \phi ],
\nonumber
\\
s_Q \bar{\phi} &= \eta,
\quad
s_Q \eta = 0,
\qquad
s_Q \chi_{\mu\nu} = \lambda_{\mu\nu},
\quad
s_Q \lambda_{\mu\nu} = 0,
\qquad
s_Q \bar{C} = B,
\quad
s_Q B = 0.
\end{align}

Next, let us adopt the preceding construction of TQCD in the covariant
Landau type gauge as a guiding principle for the gauge--fixed action of  
TYM in the non--covariant Landau type gauge \cite{17}, namely
requiring 
\begin{equation}
\label{3.2}
{W}_{\rm TYM} = s_Q {{\Psi}}_{\rm TYM}, 
\qquad
\bar{Q}_\mu {{\Psi}}_{\rm TYM} = 0,
\end{equation}
with the following modification of the gauge fermion 
(cf.,~Eqs.~(\ref{2.24}) and (\ref{2.xx}))
\begin{equation}
\label{3.3}
{{\Psi}}_{\rm TYM} = \int d^4x\, {\rm tr} \Bigr\{
\chi^{\mu\nu} F_{\mu\nu} + 
2 \bar{\phi} \partial^\mu \psi_\mu + 2 \bar{C} \partial^\mu A_\mu \Bigr\}, 
\end{equation}
which consists in choosing a linear gauge condition not only for the 
ordinary gauge symmetry but also for the topological shift symmetry. 
Then, by making use of (\ref{3.1}), one gets
\begin{align}
\label{3.4}
{W}_{\rm TYM} = \int d^4x\, {\rm tr} \Bigr\{&
( \lambda^{\mu\nu} - \{ C, \chi^{\mu\nu} \} ) F_{\mu\nu} - 
2 \chi^{\mu\nu} D_\mu \psi_\nu + 2 \eta \partial^\mu \psi_\mu
\nonumber
\\
& + 2 \bar{\phi} \partial^\mu ( D_\mu \phi + \{ C, \psi_\mu \} ) +
2 B \partial^\mu A_\mu -
2 \bar{C} \partial^\mu ( \psi_\mu - D_\mu C ) \Bigr\}. 
\end{align}
Moreover, it is easily seen that this action exhibits, besides the 
BRST symmetry (\ref{3.1}), invariance under the following linear 
vector supersymmetry \cite{17}:
\begin{align}
\label{3.5}
\bar{Q}_\mu A_\nu &= 0,
\qquad
\bar{Q}_\mu \psi_\nu = \partial_\mu A_\nu,
\nonumber
\\
\bar{Q}_\mu C &= 0,
\qquad
\bar{Q}_\mu \phi = \partial_\mu C,
\qquad
\bar{Q}_\mu \bar{\phi} = 0,
\qquad
\bar{Q}_\mu \eta = \partial_\mu \bar{\phi},
\nonumber
\\
\bar{Q}_\mu \chi_{\rho\sigma} &= 0,
\qquad
\bar{Q}_\mu \lambda_{\rho\sigma} = \partial_\mu \chi_{\rho\sigma},
\qquad
\bar{Q}_\mu \bar{C} = \partial_\mu \bar{\phi},
\qquad
\bar{Q}_\mu B = \partial_\mu \bar{C} - \partial_\mu \eta,
\end{align}
These transformations are determined by the second of
the Eqs.~(\ref{3.2}) and the requirement, together with (\ref{3.1}), 
to obey the topological superalgebra (\ref{2.36})
without central extensions.

Now, we are faced with the problem to construct an action of TQCD which, 
on the one hand, bears in mind the non--covariant gauge choice 
$\partial_\mu \psi^\mu = 0$ and, on the other hand, satisfies the 
topological superalgebra (\ref{2.36}) with the central charges
$Z$ and $\bar{Z}$. This problem amounts to look for a linear vector 
supersymmetry $\bar{Q}_\mu$ of the matter fields, too, which together with 
(\ref{2.34}), (\ref{3.1}) and (\ref{3.5}) obeys the superalgebra 
(\ref{2.36}). 
It is not difficult to convince oneself that this requirement is indeed 
fulfilled by defining the action of $\bar{Q}_\mu$ on the matter fields as
\begin{alignat}{2}
\label{3.6}
\bar{Q}_\mu \zeta^A &= 
i (\sigma_\mu)^{A \dot{B}} \bar{\beta}_{\dot{B}},
&\qquad
\bar{Q}_\mu \bar{\beta}_{\dot{A}} &= 
i \bar{m} (\sigma_\mu)_{\dot{A} B} \zeta^B,
\nonumber
\\
\bar{Q}_\mu \alpha^A &= 
- i (\sigma_\mu)^{A \dot{B}} \bar{\chi}_{\dot{B}} + \partial_\mu \zeta^A,
&\qquad
\bar{Q}_\mu \bar{\chi}_{\dot{A}} &= 
- i \bar{m} (\sigma_\mu)_{\dot{A} B} \alpha^B +
\partial_\mu \bar{\beta}_{\dot{A}},
\nonumber
\\
\bar{Q}_\mu \bar{\zeta}_A &= 
- i (\sigma_\mu)_{A \dot{B}} \bar{\alpha}^{\dot{B}},
&\qquad
\bar{Q}_\mu \bar{\alpha}^{\dot{A}} &= 
i \bar{m} (\sigma_\mu)^{\dot{A} B} \bar{\zeta}_B,
\nonumber
\\
\bar{Q}_\mu \beta_A &= 
i (\sigma_\mu)_{A \dot{B}} \chi^{\dot{B}} +
\partial_\mu \bar{\zeta}_A,
&\qquad
\bar{Q}_\mu \chi^{\dot{A}} &= 
- i \bar{m} (\sigma_\mu)^{\dot{A} B} \beta_B +
\partial_\mu \bar{\alpha}^{\dot{A}}.
\end{alignat}
The transformation rules (\ref{3.6}) are obtained from (\ref{2.35}) by simply
replacing the gauge covariant derivative through the ordinary
partial derivative. 

Since $\bar{Q}_\mu$ leaves the gauge field $A_\mu$ inert, from (\ref{3.6}) 
one infers that now the above replacement procedure simply can be repeated in
order to get the action of TQCD: 
\begin{equation}
\label{3.7}
{W}_{\rm TQCD} = {W}_{\rm TYM} + {W}_{\rm M},
\qquad
{W}_{\rm M} = s_Q {{\Psi}}_{\rm M},
\qquad
\bar{Q}_\mu {{\Psi}}_{\rm M} = 0,
\end{equation}
with the following linearized gauge fermion (see Eq.~(\ref{2.20}))
\begin{align}
\label{3.8}
{{\Psi}}_{\rm M} = - \int d^4x\, \Bigr\{&
i \bar{\alpha}^{\dot{A}} (\sigma^\mu)_{\dot{A} B} 
\partial_\mu \zeta^B -
i (\partial_\mu \bar{\zeta}_A) (\sigma^\mu)^{A \dot{B}} 
\bar{\beta}_{\dot{B}}
\\
& + \bar{m} ( \bar{\zeta}_A \alpha^A - \beta_A \zeta^A ) + 
\bar{\alpha}^{\dot{A}} \bar{\chi}_{\dot{A}} +
\chi^{\dot{A}} \bar{\beta}_{\dot{A}} \Bigr\},
\nonumber
\end{align}
and, by making use of (\ref{2.34}), the action corresponding to
the matter fields reads:
\begin{align}
\label{3.9}
{W}_{\rm M} = \int d^4x\, \Bigr\{&
i \bar{\alpha}^{\dot{A}} (\sigma^\mu)_{\dot{A} B} \partial_\mu \alpha^B +
i (\partial_\mu \beta_A) (\sigma^\mu)^{A \dot{B}} \bar{\beta}_{\dot{B}} -
2 \bar{m} \beta_A \alpha^A 
\\
& + i \bar{\alpha}^{\dot{A}} (\sigma^\mu)_{\dot{A} B} 
(\partial_\mu C) \zeta^B -
i \bar{\zeta}_A (\sigma^\mu)^{A \dot{B}} 
(\partial_\mu C) \bar{\beta}_{\dot{B}} + 
2 \bar{m} \bar{\zeta}_A ( \phi + m ) \zeta^A
\nonumber
\phantom{\frac{1}{2}}
\\
& - i \chi^{\dot{A}} (\sigma^\mu)_{\dot{A} B} \partial_\mu \zeta^B +
i (\partial_\mu \bar{\zeta}_A) (\sigma^\mu)^{A \dot{B}} 
\bar{\chi}_{\dot{B}} - 2 \chi^{\dot{A}} \bar{\chi}_{\dot{A}} - 
2 \bar{\alpha}^{\dot{A}} ( \phi + m ) \bar{\beta}_{\dot{A}} \Bigr\}.
\nonumber
\end{align}

Finally, let us notice that the ghost $C$ enters into the action (\ref{3.4}) 
only as derivative $\partial_\mu C$ as well as through the combinations 
$\eta - [ C, \bar{\phi} ]$, $\lambda^{\mu\nu} - \{ C, \chi^{\mu\nu} \}$ and 
$B - \{ C, \bar{C} \}$. These global constraints can be expressed 
by the so--called ghost equation \cite{29}, usually valid in the Landau type 
gauge. There exists still another set of global constraints, the so--called 
ghost for the ghost equation \cite{17}, being related to the non--covariant 
Landau type gauge. It expresses the fact that the ghost for the ghost $\phi$ 
enters into the action (\ref{3.4}) only as derivative $\partial_\mu \phi$ 
as well as through the combination $B - [ \bar{\phi}, \phi ]$. On the other 
hand, by inspection of the action (\ref{3.9}) one observes that only the 
first of the above--mentioned ghost equation can be imposed as global constraint,
whereas the second one leads to a non--linear breaking term.


\bigskip
\begin{flushleft}
{\large{\bf 4. Slavnov--Taylor identity and stability constraints}}
\end{flushleft}
\bigskip
In the previous section we have constructed the complete gauge--fixed action
(\ref{3.7}) of TQCD being invariant under the whole set of symmetry 
operators $s_Q$, $\bar{Q}_\mu$, $P_\mu$, $Z$ and $\bar{Z}$ obeying 
the topological superalgebra (\ref{2.36}). Our aim is now to collect all the 
symmetry properties of that action into a {\it unique} Ward identity -- 
the Slavnov--Taylor identity. This could be achieved in the conventional 
Batalin--Vilkovisky (BV) approach \cite{30} where for every field a 
corresponding antifield with opposite Grassmann parity is introduced and then
the symmetry properties are compactly formulated by some master equation. 
Here, however, we adopt the completely equivalent method of 
Ref.~\cite{31}, where the antifields (sources) 
are introduced only for the fields belonging to the minimal sector, and
where the symmetry operators $s_Q$, $\bar{Q}_\mu$, $P_\mu$, $Z$ and 
$\bar{Z}$ are collected into a unique nilpotent BRST operator $s_T$ by 
associating to each of the generators $\bar{Q}_\mu$, $P_\mu$, $Z$ and 
$\bar{Z}$ a global ghost $\rho^\mu$, $\xi^\mu$, $\bar{\xi}$ 
and $\xi$, respectively. 

In this context let us recall that identifying the $R$--charge with the 
ghost number has the meaning of setting the global ghost $\rho$ associated 
to the generator $Q$ equal to one, i.e., $\rho = 1$. This already has
been done by defining the operator $s_Q$ according to $s_Q = s + Q$ and 
regarding $Q$ as a BRST--like operator. Then, the aforementioned  
Slavnov--Taylor identity can be derived in the usual manner by coupling 
the antifields 
to the non--linear parts of field transformations generated by $s_Q$. 
In addition, the action (\ref{3.7}) obeys a set of gauge--fixing and
antighost conditions as well as global constraints being related to the 
non--covariant Landau type gauge.


\bigskip
\begin{flushleft}
{4.1. {\it Introduction of antifields}}
\end{flushleft}
\bigskip
Let us begin by introducing a set of global ghosts $\rho^\mu$, $\xi^\mu$,
$\bar{\xi}$ and $\xi$, associated, respectively, to the generators 
$\bar{Q}_\mu$, $P_\mu$, $Z$ and $\bar{Z}$, and defining, in this way, the 
total BRST operator
\begin{gather}
\label{4.1}
s_T = s_Q + \rho^\mu \bar{Q}_\mu -i \xi^\mu P_\mu + 
\bar{\xi} Z + \xi \bar{Z}, 
\\
{\rm with}
\qquad
s_T \rho^\mu = 0,
\qquad
s_T \xi^\mu = - \rho^\mu,
\qquad
s_T \bar{\xi} = 1,
\qquad
s_T \xi = - \rho^\mu \rho_\mu.
\nonumber
\end{gather}
Here, the transformation rules of the global ghosts are chosen in such a way 
that $s_T$ is strictly nilpotent. Then, it holds 
\begin{equation*}
s_T {W}_{\rm TQCD} = 0,
\qquad
\{ s_T, s_T \} = 0,
\end{equation*}
where all the relevant features of the superalgebra (\ref{2.36}) are 
now encoded in the nilpotency of the unique BRST operator $s_T$. 
The properties of the global ghosts are displayed in the following
Table 4.
\bigskip

\begin{equation*}
\hbox{
\setlength{\extrarowheight}{3pt}
\begin{tabular}{|c|c|c|c|c|c|}
\hline
& $\rho^\mu$ & $\xi^\mu$ & $\bar{\xi}$ & $\xi$  
\\ 
\hline
ghost number & 2 & 1 & --1 & 3  
\\ 
\hline
mass dimension & 0 & --1/2 & --1/2 & --1/2 
\\ 
\hline
Grassmann parity & even & odd & odd & odd
\\
\hline
\end{tabular}}
\end{equation*}
\bigskip

Let us now introduce the antifields of the minimal sector,
$V_T^* = \{ A_\mu^*, \psi_\mu^*, C^*, \phi^* \}$, transforming according to 
\begin{equation}
\label{4.2}
s_Q \psi_\mu^* = A_\mu^*,
\qquad
s_Q A_\mu^* = 0,
\qquad
s_Q \phi^* = C^*, 
\qquad
s_Q C^* = 0
\end{equation}
and
\begin{equation}
\label{4.3}
\bar{Q}_\mu \psi_\nu^* = 0,
\qquad
\bar{Q}_\mu A_\nu^* = \partial_\mu \psi_\nu^*,
\qquad
\bar{Q}_\mu \phi^* = 0, 
\qquad
\bar{Q}_\mu C^* = \partial_\mu \phi^*,
\end{equation}
so that the superalgebra (\ref{2.36}) is satisfied. In particular, from 
Eqs. (\ref{4.2}) it is obvious that the antifields are grouped in 
BRST--doublets.

Next, we extend the action (\ref{3.4}) by adding a pure BRST invariant term 
$s_Q \Upsilon_{\rm T}$, 
\begin{equation}
\label{4.4}
S_{\rm TYM} = {W}_{\rm TYM} + s_Q \Upsilon_{\rm T},
\end{equation}
with
\begin{align}
\Upsilon_{\rm T} = - \int d^4x\, {\rm tr} \Bigr\{&
A_\mu^* A^\mu + \psi_\mu^* \psi^\mu - C^* C + \phi^* \phi \Bigr\}
\nonumber
\\
\label{4.5}
s_Q \Upsilon_{\rm T} = \int d^4x\, {\rm tr} \Bigr\{&
A_\mu^* ( s_Q A^\mu - \psi^\mu ) - \psi_\mu^* (s_Q \psi^\mu) +
C^* ( s_Q C - \phi ) + \phi^* (s_Q \phi) \Bigr\}.
\end{align}
Notice, that $\Upsilon_{\rm T}$, by virtue of (\ref{3.5}), does not violate 
the vector supersymmetry $\bar{Q}_\mu$, i.e., 
it holds $\bar{Q}_\mu \Upsilon_{\rm T} = 0$. Then, by 
making use of (\ref{3.1}), for the antifield dependent terms we obtain
\begin{equation}
\label{4.6}
s_Q \Upsilon_{\rm T} = \int d^4x\, {\rm tr} \Bigr\{
- A_\mu^* D^\mu C - \psi_\mu^* ( D^\mu \phi + \{ C, \psi^\mu \} ) +
C^* C^2 + \phi^* [ C, \phi ] \Bigr\}.
\end{equation}

Here, it is worthwile do draw the attention to a particular feature of the 
operator $s_Q$. First, since the antifields in the transformation law 
(\ref{4.2}) appear in BRST--doublets they do not contribute to the BRST 
cohomology \cite{32}. Second, the antifields in (\ref{4.4}) do not couple to 
the linear parts $\psi_\mu$ and $\phi$ of the field transformations 
$s_Q A^\mu$ and $s_Q C$ (cf.,~Eqs.~(\ref{3.1})). 

\smallskip
{\small
Furthermore, as a useful hint we remark that also
the fields of the minimal sector could be cast into BRST--doublets 
as well, namely
\begin{equation*}
s_Q A_\mu = \psi_\mu,
\qquad
s_Q \psi_\mu = 0,
\qquad
s_Q C = \phi,
\qquad
s_Q \phi = 0.
\end{equation*}
This might be achieved by redefining $\psi_\mu$ and $\phi$ according to  
the replacements 
\begin{eqnarray}
\psi_\mu \rightarrow \psi_\mu + D_\mu C
\qquad 
\hbox{and} 
\qquad
\phi \rightarrow \phi - C^2.
\nonumber
\end{eqnarray}
For that reason, one expects that the cohomology of the operator 
$s_Q$ is completely trivial \cite{5,23}. However, needless to say, 
such redefinitions can be performed only at the lowest order 
of perturbation theory.}
\smallskip

Finally, let us display the properties of the antifields of the gauge 
multiplet in Table 5.
\bigskip
\begin{equation*}
\hbox{
\setlength{\extrarowheight}{3pt}
\begin{tabular}{|c|c|c|c|c|}
\hline
& $A_\mu^*$ & $\psi_\mu^*$ & $C^*$ & $\phi^*$ 
\\ 
\hline
ghost number & --1 & --2 & --2 & --3  
\\ 
\hline
mass dimension & 3 & 5/2 & 5/2 & 3 
\\ 
\hline
scale dimension & 3 & 3 & 4 & 4  
\\ 
\hline
$Z$--, $\bar{Z}$--charge & 0 & 0 & 0 & 0  
\\ 
\hline
Grassmann parity & odd & even & even & odd 
\\
\hline
\end{tabular}}
\end{equation*}
\bigskip

Proceeding in the same manner as before, let us introduce for each 
matter field a corresponding antifield, $Y^*_T = \{ 
\alpha_A^*, \bar{\beta}^{\dot{A} *}, \zeta_A^*, \bar{\chi}^{\dot{A} *} \}$ 
and $\bar{Y}^*_T = \{ 
\beta^{A *}, \bar{\alpha}_{\dot{A}}^*, \bar{\zeta}^{A *}, 
\chi_{\dot{A}}^* \}$, 
transforming according to 
\begin{alignat}{4}
\label{4.7}
s_Q \alpha_A^* &= \zeta_A^*,
&\qquad
s_Q \zeta_A^* &= - m \alpha_A^*,
&\qquad
s_Q \bar{\chi}^{\dot{A} *} &= \bar{\beta}^{\dot{A} *}, 
&\qquad
s_Q \bar{\beta}^{\dot{A} *} &= - m \bar{\chi}^{\dot{A} *},
\nonumber
\\
s_Q \beta^{A *} &= \bar{\zeta}^{A *},
&\qquad
s_Q \bar{\zeta}^{A *} &= m \beta^{A *},
&\qquad
s_Q \chi_{\dot{A}}^* &= \bar{\alpha}_{\dot{A}}^*,
&\qquad 
s_Q \bar{\alpha}_{\dot{A}}^* &= m \chi_{\dot{A}}^*,
\end{alignat}
and
\begin{alignat}{2}
\label{4.8}
\bar{Q}_\mu \bar{\chi}^{\dot{A} *} &= 
i (\sigma_\mu)^{\dot{A} B} \alpha_B^*,
&\qquad
\bar{Q}_\mu \alpha_A^* &= 
- i \bar{m} (\sigma_\mu)_{A \dot{B}} \bar{\chi}^{\dot{B} *},
\nonumber
\\
\bar{Q}_\mu \bar{\beta}^{\dot{A} *} &= 
- i (\sigma_\mu)^{\dot{A} B} \zeta_B^* +
\partial_\mu \bar{\chi}^{\dot{A} *},
&\qquad
\bar{Q}_\mu \zeta_A^* &= 
i \bar{m} (\sigma_\mu)_{A \dot{B}} \bar{\beta}^{\dot{B} *} +
\partial_\mu \alpha_A^*,
\nonumber
\\
\bar{Q}_\mu \chi_{\dot{A}}^* &= 
i (\sigma_\mu)_{\dot{A} B} \beta^{B *},
&\qquad
\bar{Q}_\mu \beta^{A *} &= 
i \bar{m} (\sigma_\mu)^{A \dot{B}} \chi_{\dot{B}}^*,
\nonumber
\\
\bar{Q}_\mu \bar{\alpha}_{\dot{A}}^* &= 
- i (\sigma_\mu)_{\dot{A} B} \bar{\zeta}^{B *} + 
\partial_\mu \chi_{\dot{A}}^*,
&\qquad
\bar{Q}_\mu \bar{\zeta}^{A *} &= 
- i \bar{m} (\sigma_\mu)^{A \dot{B}} \bar{\alpha}_{\dot{B}}^* +
\partial_\mu \beta^{A *},
\end{alignat}
which together with the central charge transformations
\begin{align}
\label{4.9}
Z V^*_T &= 0,
\qquad 
Z Y_T^* = -m Y_T^*,
\qquad
Z \bar{Y}_T^* = m \bar{Y}_T^*,
\nonumber
\\
\bar{Z} V^*_T &= 0,
\qquad 
\bar{Z} Y_T^* = - \bar{m} Y_T^*,
\qquad
Z\bar{ }\bar{Y}_T^* = \bar{m} \bar{Y}_T^*,
\end{align}
obey the superalgebra (\ref{2.36}). 

Next, let us add also to the matter action 
(\ref{3.9}) a pure BRST invariant term $s_Q \Upsilon_{\rm M}$,
\begin{align}
\label{4.10}
S_{\rm M} = {W}_{\rm M} + s_Q \Upsilon_{\rm M},
\end{align}
{with}
\begin{align}
\Upsilon_{\rm M} = - \int d^4x\, \Bigr\{&
\alpha_A^* \alpha^A + \zeta_A^* \zeta^A + 
\bar{\chi}^{\dot{A} *} \bar{\chi}_{\dot{A}} - 
\bar{\beta}^{\dot{A} *} \bar{\beta}_{\dot{A}}
\nonumber
\\
& + \beta^{A *} \beta_A + \bar{\zeta}^{A *} \bar{\zeta}_A +
\chi_{\dot{A}}^* \chi^{\dot{A}} - 
\bar{\alpha}_{\dot{A}}^* \bar{\alpha}^{\dot{A}} \Bigr\}
\nonumber
\\
\label{4.11}
s_Q \Upsilon_{\rm M} = \int d^4x\, \Bigr\{&
- \alpha_A^* ( s_Q \alpha^A - m \zeta^A ) + 
\zeta_A^* ( s_Q \zeta^A - \alpha^A )
\nonumber
\\
& + \bar{\chi}^{\dot{A} *} ( s_Q \bar{\chi}_{\dot{A}} - 
m \bar{\beta}_{\dot{A}} )  + \bar{\beta}^{\dot{A} *} 
( s_Q \bar{\beta}_{\dot{A}} - \bar{\chi}_{\dot{A}} )
\nonumber
\\
\phantom{\frac{1}{2}}
& - \beta^{A *} ( s_Q \beta_A + m \bar{\zeta}_A ) + 
\bar{\zeta}^{A *} ( s_Q \bar{\zeta}_A - \beta_A )
\nonumber
\\
& + \chi_{\dot{A}}^* ( s_Q \chi^{\dot{A}} + m \alpha^{\dot{A}} ) + 
\bar{\alpha}_{\dot{A}}^* ( s_Q \bar{\alpha}^{\dot{A}} - \chi^{\dot{A}} ) 
\Bigr\}.
\end{align}
As before, by virtue of (\ref{3.6}), it can be verified that 
$\Upsilon_{\rm M}$ does not spoil the vector supersymmetry $\bar{Q}_\mu$, 
i.e., it holds $\bar{Q}_\mu \Upsilon_{\rm M} = 0$. Thus, by making use of 
(\ref{2.34}), for the antifield dependent terms we get
\begin{align}
\label{4.12}
s_Q \Upsilon_{\rm M} = \int d^4x\, \Bigr\{&
\alpha_A^* ( \phi \zeta^A - C \alpha^A ) + 
\zeta_A^* C \zeta^A - \bar{\chi}^{\dot{A} *} ( 
\phi \bar{\beta}_{\dot{A}} - C \bar{\chi}_{\dot{A}} ) +
\bar{\beta}^{\dot{A} *} C \bar{\beta}_{\dot{A}}
\nonumber
\\
& - \beta^{A *} ( \bar{\zeta}_A \phi + \beta_A C ) - 
\bar{\zeta}^{ A *} \bar{\zeta}_A C +
\chi_{\dot{A}}^* ( \bar{\alpha}^{\dot{A}} \phi - \chi^{\dot{A}} C ) +
\bar{\alpha}_{\dot{A}}^* \bar{\alpha}^{\dot{A}} C \Bigr\}.
\end{align}

\smallskip
{\small
Again, here one observes that when carrying out in (\ref{2.34}) 
the replacements 
\begin{eqnarray}
\alpha^A \rightarrow \alpha^A - C \zeta^A,
\qquad 
\beta^A \rightarrow \beta^A + \bar{\zeta}_A C,
\qquad
\bar{\chi}_{\dot{A}} \rightarrow \bar{\chi}_{\dot{A}} - 
C \bar{\beta}_{\dot{A}},
\qquad 
\chi^{\dot{A}} \rightarrow \chi^{\dot{A}} -
\bar{\alpha}^{\dot{A}} C,
\nonumber
\end{eqnarray}
a doublet--like structure of the matter fields would follow, 
\begin{align*}
s_Q \zeta^A &= \alpha^A,
\qquad
s_Q \alpha^A =  m \zeta^A,
\qquad
s_Q \bar{\beta}_{\dot{A}} = \bar{\chi}_{\dot{A}},
\qquad
s_Q \bar{\chi}_{\dot{A}} =  m \bar{\beta}_{\dot{A}},
\\
s_Q \bar{\zeta}_A &= \beta_A,
\qquad
s_Q \beta_A = - m \bar{\zeta}_A,
\qquad
s_Q \bar{\alpha}^{\dot{A}} = \chi^{\dot{A}},
\qquad
s_Q \chi^{\dot{A}} = - m \alpha^{\dot{A}},
\end{align*}
showing that the cohomology of the operator $s_Q$, when acting on the 
space of integrated local polynomials of the fields and antifields with 
vanishing central charge, would be trivial, too. }
\smallskip

The properties of the matter--antifields are displayed in Table 6.
\bigskip
\begin{equation*}
\hbox{
\setlength{\extrarowheight}{3pt}
\begin{tabular}{|c|c|c|c|c|c|c|c|c|}
\hline
& $\alpha_A^*$ & $\beta^{A *}$ & 
$\bar{\alpha}_{\dot{A}}^*$ & $\bar{\beta}^{\dot{A} *}$ 
& $\zeta_A^*$ & $\zeta^{A *}$ 
& $\chi_{\dot{A}}^*$ & $\bar{\chi}^{\dot{A} *}$  
\\ 
\hline
ghost number & --2 & --2 & --2 & --2 & --1 & --1 & --1 & --1 
\\ 
\hline
mass dimension & 5/2 & 5/2 & 5/2 & 5/2 & 3 & 3 & 2 & 2
\\ 
\hline
scale dimension & 3 & 3 & 2 & 2 & 3 & 3 & 1 & 1
\\ 
\hline
$Z$--, $\bar{Z}$--charge & 1 & --1 & --1 & 1 & 
                           1 & --1 & --1 & 1 
\\ 
\hline
Grassmann parity & odd & odd & odd & odd & even & even & even & even 
\\
\hline
\end{tabular}}
\end{equation*}
\bigskip

Finally, putting together (\ref{4.4}) and (\ref{4.10}) we get the 
extended antifield-dependent action of TQCD,
\begin{equation}
\label{4.13}
S_{\rm TQCD} = S_{\rm TYM} + S_{\rm M}, 
\qquad
s_T S_{\rm TQCD} = 0,
\end{equation}
where any of its symmetry properties is encoded in the single
operator $s_T$.


\bigskip
\begin{flushleft}
{4.2. {\it Slavnov--Taylor identity}}
\end{flushleft}
\bigskip
We translate now the symmetry property (\ref{4.13}) into the 
Slavnov--Taylor identity,
\begin{equation}
\label{4.14}
\mathbf{S}_T(S) = 0,
\end{equation}
of the classical action $S_{\rm TQCD}$ which, for notational
simplicity, has been denoted by $S$.
Here, $\mathbf{S}_T(S)$ displays the following expansion with respect to
the global ghosts,
\begin{equation}
\label{4.15}
\mathbf{S}_T(S) \equiv \mathbf{S}_Q(S) + \rho^\mu \mathbf{\bar{Q}}_\mu S - 
i \xi^\mu \mathbf{P}_\mu S + \bar{\xi} \mathbf{Z} S + \xi \mathbf{\bar{Z}} S - 
\rho^\mu \frac{\partial S}{\partial \xi^\mu} + 
\frac{\partial S}{\partial \bar{\xi}} -
\rho^\mu \rho_\mu \frac{\partial S}{\partial \xi},
\end{equation}
where $\mathbf{\bar{Q}}_\mu$, $\mathbf{P}_\mu$, $\mathbf{Z}$ and 
$\mathbf{\bar{Z}}$ denote the Ward operators of vector supersymmetry 
transformations, space--time translations and central charge transformations 
in the space of fields and antifields, respectively. Furthermore, we 
represent the nilpotent BRST operator $s_T$ by a {\em linear} nilpotent 
functional differential operator:
\begin{equation}
\label{4.16}
\mathbf{S}_T = \mathbf{S}_Q + \rho^\mu \mathbf{\bar{Q}}_\mu - i 
\xi^\mu \mathbf{P}_\mu + \bar{\xi} \mathbf{Z} + \xi \mathbf{\bar{Z}} - 
\rho^\mu \frac{\partial }{\partial \xi^\mu} + 
\frac{\partial }{\partial \bar{\xi}} -
\rho^\mu \rho_\mu \frac{\partial }{\partial \xi},
\end{equation}
where $\mathbf{\bar{Q}}_\mu$, $\mathbf{P}_\mu$, $\mathbf{Z}$ and 
$\mathbf{\bar{Z}}$, 
together with $\mathbf{S}_Q$, are required to obey the topological superalgebra 
\begin{equation}
\label{4.17}
\{ \mathbf{S}_Q, \mathbf{S}_Q \} = - 2 \mathbf{Z},
\qquad
\{ \mathbf{S}_Q, \mathbf{\bar{Q}}_\mu \} = - i \mathbf{P}_\mu,
\qquad
\{ \mathbf{\bar{Q}}_\mu, \mathbf{\bar{Q}}_\nu \} = - 
2 \delta_{\mu\nu} \mathbf{\bar{Z}}.
\end{equation}
From Eq.~(\ref{4.17}) it follows that $\mathbf{S}_Q$ 
becomes a nilpotent operator when acting on the space of integrated local
polynomials with vanishing central charge (and being independent of
the global ghosts). 
Finally, we obtain $\{ \mathbf{S}_T, \mathbf{S}_T \} = 0$.

In (\ref{4.15}) the operator $\mathbf{S}_Q(S)$ lumps 
together both the linear and non--linear parts of the BRST transformations, 
the latter ones being expressed by derivatives with respect to the antifields 
(see Eqs. (\ref{3.1}), (\ref{4.2}), (\ref{4.5}), (\ref{4.7}) and 
(\ref{4.11})). It takes the form 
\begin{align}
\mathbf{S}_Q(S) = \int d^4x\, {\rm tr} \Bigr\{&
\Bigr( \frac{\delta S}{\delta A_\mu^*} + \psi^\mu \Bigr)
\frac{\delta S}{\delta A^\mu} + 
\Bigr( \frac{\delta S}{\delta C^*} + \phi \Bigr)
\frac{\delta S}{\delta C}
\\
& - \Bigr( \frac{\delta S}{\delta \psi^\mu} - A_\mu^* \Bigr)
\frac{\delta S}{\delta \psi_\mu^*} +
\Bigr( \frac{\delta S}{\delta \phi} + C^* \Bigr) 
\frac{\delta S}{\delta \phi^*} + 
B \frac{\delta S}{\delta \bar{C}} +
\eta \frac{\delta S}{\delta \bar{\phi}} + 
\frac{1}{2} \lambda_{\mu\nu} \frac{\delta S}{\delta \chi_{\mu\nu}} 
\Bigr\}
\nonumber
\\
+ \int d^4x\, \Bigr\{&
\Bigr( \frac{\delta S}{\delta \zeta_A^*} + \alpha^A \Bigr)
\frac{\delta S}{\delta \zeta^A} +
m \zeta^A \frac{\delta S}{\delta \alpha^A} +
\Bigr( \frac{\delta S}{\delta \bar{\beta}^{\dot{A} *}} + 
\bar{\chi}_{\dot{A}} \Bigr) \frac{\delta S}{\delta \bar{\beta}_{\dot{A}}} +
m \bar{\beta}_{\dot{A}} \frac{\delta S}{\delta \bar{\chi}_{\dot{A}}}
\nonumber
\\
& + \Bigr( \frac{\delta S}{\delta \alpha^A} + \zeta_A^* \Bigr)
\frac{\delta S}{\delta \alpha_A^*} - 
m \alpha_A^* \frac{\delta S}{\delta \zeta_A^*} -
\Bigr( \frac{\delta S}{\delta \bar{\chi}_{\dot{A}}} - 
\bar{\beta}^{\dot{A} *} \Bigr) 
\frac{\delta S}{\delta \bar{\chi}^{\dot{A} *}} -
m \bar{\chi}^{\dot{A} *} \frac{\delta S}{\delta \bar{\beta}^{\dot{A} *}}
\nonumber
\\
& + \Bigr( \frac{\delta S}{\delta \bar{\zeta}^{A *}} + \beta_A \Bigr)
\frac{\delta S}{\delta \bar{\zeta}_A} -
m \bar{\zeta}_A \frac{\delta S}{\delta \beta_A} +
\Bigr( \frac{\delta S}{\delta \bar{\alpha}_{\dot{A}}^*} + 
\chi^{\dot{A}} \Bigr) \frac{\delta S}{\delta \bar{\alpha}^{\dot{A}}} -
m \bar{\alpha}^{\dot{A}} \frac{\delta S}{\delta \chi^{\dot{A}}}
\nonumber
\\
& + \Bigr( \frac{\delta S}{\delta \beta_A} + \bar{\zeta}^{A *} \Bigr)
\frac{\delta S}{\delta \beta^{A *}} +
m \beta^{A *} \frac{\delta S}{\delta \bar{\zeta}^{A *}} -
\Bigr( \frac{\delta S}{\delta \chi^{\dot{A}}} - 
\bar{\alpha}_{\dot{A}}^* \Bigr)
\frac{\delta S}{\delta \chi_{\dot{A}}^*} +
m \chi_{\dot{A}}^* \frac{\delta S}{\delta \bar{\alpha}_{\dot{A}}^*} 
\Bigr\}.
\nonumber
\end{align}
From $\mathbf{S}_Q(S)$ one reads off the {\em linearized} Slavnov--Taylor 
operator
\begin{align}
\label{4.x}
\mathbf{S}_Q = \int d^4x\, {\rm tr} \Bigr\{&
\Bigr( \frac{\delta S}{\delta A_\mu^*} + \psi^\mu \Bigr)
\frac{\delta }{\delta A^\mu} + 
\frac{\delta S}{\delta A^\mu} \frac{\delta }{\delta A_\mu^*} + 
\Bigr( \frac{\delta S}{\delta C^*} + \phi \Bigr)
\frac{\delta }{\delta C} + 
\frac{\delta S}{\delta C} \frac{\delta }{\delta C^*} 
\nonumber
\\
& - \Bigr( \frac{\delta S}{\delta \psi^\mu} - A_\mu^* \Bigr)
\frac{\delta }{\delta \psi_\mu^*} -
\frac{\delta S}{\delta \psi_\mu^*} \frac{\delta }{\delta \psi^\mu} +
\Bigr( \frac{\delta S}{\delta \phi} + C^* \Bigr) 
\frac{\delta }{\delta \phi^*} +
\frac{\delta S}{\delta \phi^*} \frac{\delta }{\delta \phi}
\nonumber
\\
&\! + B \frac{\delta }{\delta \bar{C}} +
\eta \frac{\delta }{\delta \bar{\phi}} + 
\frac{1}{2} \lambda_{\mu\nu} \frac{\delta }{\delta \chi_{\mu\nu}} 
\Bigr\}
\nonumber
\\
+ \int d^4x\, \Bigr\{&
\Bigr( \frac{\delta S}{\delta \zeta_A^*} + \alpha^A \Bigr)
\frac{\delta }{\delta \zeta^A} -
\frac{\delta S}{\delta \zeta^A} \frac{\delta }{\delta \zeta_A^*} +
\Bigr( \frac{\delta S}{\delta \bar{\beta}^{\dot{A} *}} + 
\bar{\chi}_{\dot{A}} \Bigr) \frac{\delta }{\delta \bar{\beta}_{\dot{A}}} -
\frac{\delta S}{\delta \bar{\beta}_{\dot{A}}}
\frac{\delta }{\delta \bar{\beta}^{\dot{A} *}}
\nonumber
\\
& + \Bigr( \frac{\delta S}{\delta \alpha^A} + \zeta_A^* \Bigr)
\frac{\delta }{\delta \alpha_A^*} -
\frac{\delta S}{\delta \alpha_A^*} \frac{\delta }{\delta \alpha^A} - 
\Bigr( \frac{\delta S}{\delta \bar{\chi}_{\dot{A}}} - 
\bar{\beta}^{\dot{A} *} \Bigr) 
\frac{\delta }{\delta \bar{\chi}^{\dot{A} *}} +
\frac{\delta S}{\delta \bar{\chi}^{\dot{A} *}}
\frac{\delta }{\delta \bar{\chi}_{\dot{A}}}
\nonumber
\\
& + m \zeta^A \frac{\delta }{\delta \alpha^A} +
m \bar{\beta}_{\dot{A}} \frac{\delta }{\delta \bar{\chi}_{\dot{A}}} - 
m \alpha_A^* \frac{\delta }{\delta \zeta_A^*} -
m \bar{\chi}^{\dot{A} *} \frac{\delta }{\delta \bar{\beta}^{\dot{A} *}}
\nonumber
\\
& + \Bigr( \frac{\delta S}{\delta \bar{\zeta}^{A *}} + \beta_A \Bigr)
\frac{\delta }{\delta \bar{\zeta}_A} -
\frac{\delta S}{\delta \bar{\zeta}_A} 
\frac{\delta }{\delta \bar{\zeta}^{A *}} + 
\Bigr( \frac{\delta S}{\delta \bar{\alpha}_{\dot{A}}^*} + 
\chi^{\dot{A}} \Bigr) \frac{\delta }{\delta \bar{\alpha}^{\dot{A}}} -
\frac{\delta S}{\delta \bar{\alpha}^{\dot{A}}}
\frac{\delta }{\delta \bar{\alpha}_{\dot{A}}^*}
\nonumber
\\
& + \Bigr( \frac{\delta S}{\delta \beta_A} + \bar{\zeta}^{A *} \Bigr)
\frac{\delta }{\delta \beta^{A *}} -
\frac{\delta S}{\delta \beta^{A *}} \frac{\delta }{\delta \beta_A} - 
\Bigr( \frac{\delta S}{\delta \chi^{\dot{A}}} - 
\bar{\alpha}_{\dot{A}}^* \Bigr)
\frac{\delta }{\delta \chi_{\dot{A}}^*} +
\frac{\delta S}{\delta \chi_{\dot{A}}^*}
\frac{\delta }{\delta \chi^{\dot{A}}} 
\nonumber
\\
& - m \bar{\zeta}_A \frac{\delta }{\delta \beta_A} -
m \bar{\alpha}^{\dot{A}} \frac{\delta }{\delta \chi^{\dot{A}}} +
m \beta^{A *} \frac{\delta }{\delta \bar{\zeta}^{A *}} +
m \chi_{\dot{A}}^* \frac{\delta }{\delta \bar{\alpha}_{\dot{A}}^*} 
\Bigr\}.
\end{align}

If the action $S$ is a solution of $\mathbf{S}_Q(S) = 0$ then, 
by a tedious but straightforward calculation, one can show that 
$\mathbf{S}_Q$ obeys the relation 
$\{ \mathbf{S}_Q, \mathbf{S}_Q \} = - 2 \mathbf{Z}$, 
where the central charge operators $\mathbf{Z}$ and $\mathbf{\bar Z}$
are given by (cf.,~Eqs.~(\ref{2.16}) and (\ref{4.9})) 
\begin{align*}
\mathbf{Z}/m =  \mathbf{\bar Z}/{\bar m} =
\int d^4x\, \Bigr\{&
- \alpha^A \frac{\delta}{\delta \alpha^A} -
\bar{\beta}_{\dot{A}} \frac{\delta}{\delta \bar{\beta}_{\dot{A}}} - 
\zeta^A \frac{\delta}{\delta \zeta^A} -
\bar{\chi}_{\dot{A}} \frac{\delta}{\delta \bar{\chi}_{\dot{A}}}
\\
& + \beta_A \frac{\delta}{\delta \beta_A} + 
\bar{\alpha}^{\dot{A}} \frac{\delta}{\delta \bar{\alpha}^{\dot{A}}} + 
\bar{\zeta}_A \frac{\delta}{\delta \bar{\zeta}_A} +
\chi^{\dot{A}} \frac{\delta}{\delta \chi^{\dot{A}}}
\\
& + \alpha_A^* \frac{\delta}{\delta \alpha_A^*} +
\bar{\beta}^{\dot{A} *} \frac{\delta}{\delta \bar{\beta}^{\dot{A} *}} + 
\zeta_A^* \frac{\delta}{\delta \zeta_A^*} +
\bar{\chi}^{\dot{A} *} \frac{\delta}{\delta \bar{\chi}^{\dot{A} *}}
\\
& - \beta^{A *} \frac{\delta}{\delta \beta^{A *}} -
\bar{\alpha}_{\dot{A}}^* \frac{\delta}{\delta \bar{\alpha}_{\dot{A}}^*} - 
\bar{\zeta}^{A *} \frac{\delta}{\delta \bar{\zeta}^{A *}} -
\chi_{\dot{A}}^* \frac{\delta}{\delta \chi_{\dot{A}}^*} \Bigr\}.
\end{align*}
Furthermore, if $S$ is also a solution of $\mathbf{\bar{Q}}_\mu S = 0$, 
where $\mathbf{\bar{Q}}_\mu$ is given by (cf.,~Eqs.~(\ref{3.5}), (\ref{3.6}),
(\ref{4.3}) and (\ref{4.8}))
\begin{align*}
\mathbf{\bar{Q}}_\mu = \int d^4x\, {\rm tr} \Bigr\{&
\partial_\mu A_\nu \frac{\delta}{\delta \psi_\nu} +
\partial_\mu \psi_\nu^* \frac{\delta}{\delta A_\nu^*} +
\partial_\mu C \frac{\delta}{\delta \phi} +
\partial_\mu \phi^* \frac{\delta}{\delta C^*}
\\
& + \partial_\mu \bar{\phi} \frac{\delta}{\delta \bar{C}} +
(\partial_\mu \bar{C} - \partial_\mu \eta ) \frac{\delta}{\delta B} +
\partial_\mu \bar{\phi} \frac{\delta}{\delta \eta} +
\frac{1}{2} \partial_\mu \chi_{\rho\sigma} 
\frac{\delta}{\delta \lambda_{\rho\sigma}} \Bigr\}
\\
+ \int d^4x\, \Bigr\{&
( \partial_\mu \bar{\zeta}_A +
i (\sigma_\mu)_{A \dot{B}} \chi^{\dot{B}} )
\frac{\delta}{\delta \beta_A} +
( \partial_\mu \zeta^A - i (\sigma_\mu)^{A \dot{B}} \bar{\chi}_{\dot{B}} )
\frac{\delta}{\delta \alpha^A}
\\
& + ( \partial_\mu \beta^{A *} -
i \bar{m} (\sigma_\mu)^{A \dot{B}} \bar{\alpha}_{\dot{B}}^* )
\frac{\delta}{\delta \bar{\zeta}^{A *}} +
( \partial_\mu \alpha_A^* +
i \bar{m} (\sigma_\mu)_{A \dot{B}} \bar{\beta}^{\dot{B} *} )
\frac{\delta}{\delta \zeta_A^*}
\\
& + i (\sigma_\mu)^{A \dot{B}} \bar{\beta}_{\dot{B}} 
\frac{\delta}{\delta \zeta^A} +
i \bar{m} (\sigma_\mu)_{\dot{A} B} \zeta^B 
\frac{\delta}{\delta \bar{\beta}_{\dot{A}}} + 
( \partial_\mu \bar{\alpha}^{\dot{A}} -
i \bar{m} (\sigma_\mu)^{\dot{A} B} \beta_B )
\frac{\delta}{\delta \chi^{\dot{A}}}
\\
& - i (\sigma_\mu)_{A \dot{B}} \bar{\alpha}^{\dot{B}}
\frac{\delta}{\delta \bar{\zeta}_A} +
i \bar{m} (\sigma_\mu)^{\dot{A} B} \bar{\zeta}_B
\frac{\delta}{\delta \bar{\alpha}^{\dot{A}}} + 
( \partial_\mu \bar{\beta}_{\dot{A}} - 
i \bar{m} (\sigma_\mu)_{\dot{A} B} \alpha^B ) 
\frac{\delta}{\delta \bar{\chi}_{\dot{A}}}
\\
& + i (\sigma_\mu)_{\dot{A} B} \beta^{B *}
\frac{\delta}{\delta \chi_{\dot{A}}^*} +
i \bar{m} (\sigma_\mu)^{A \dot{B}} \chi_{\dot{B}}^*
\frac{\delta}{\delta \beta^{A *}} +
( \partial_\mu \chi_{\dot{A}}^* - 
i (\sigma_\mu)_{\dot{A} B} \bar{\zeta}^{B *} )
\frac{\delta}{\delta \bar{\alpha}_{\dot{A}}^*} 
\\
& + i (\sigma_\mu)^{\dot{A} B} \alpha_B^*
\frac{\delta}{\delta \bar{\chi}^{\dot{A} *}} -
i \bar{m} (\sigma_\mu)_{A \dot{B}} \bar{\chi}^{\dot{B} *}
\frac{\delta}{\delta \alpha_A^*} + 
( \partial_\mu \bar{\chi}^{\dot{A} *} -
i (\sigma_\mu)^{\dot{A} B} \zeta_B^* )
\frac{\delta}{\delta \bar{\beta}^{\dot{A} *}} \Bigr\},
\end{align*}
then one can 
verify that the vector supersymmetry $\mathbf{\bar{Q}}_\mu$ 
allows to decompose the translation operator $\mathbf{P}_\mu$ as 
$\{ \mathbf{\bar{Q}}_\mu, \mathbf{S}_Q \} = - i \mathbf{P}_\mu$. 
Moreover, one can establish, in accordance with the superalgebra (\ref{4.17}),
that it holds $\{ \mathbf{\bar{Q}}_\mu, \mathbf{\bar{Q}}_\nu \} = 
- 2 \delta_{\mu\nu} \mathbf{\bar{Z}}$.
  

\bigskip
\begin{flushleft}
{4.3. {\it Equations of motion and global constraints}}
\end{flushleft}
\bigskip
Besides obeying the Slavnov--Taylor identity (\ref{4.14}), the complete 
action $S$ turns out to be 
characterized by further constraints, namely both Landau gauge--fixing 
conditions (see Eqs.~(\ref{3.4}) and (\ref{4.6}))
\begin{equation}
\label{4.18}
\frac{\delta S}{\delta B} = 2 \partial_\mu A^\mu,
\qquad
\frac{\delta S}{\delta \eta} = 2 \partial_\mu \psi^\mu,
\end{equation}
and the corresponding antighost equations of motion
\begin{equation}
\label{4.19}
\frac{\delta S}{\delta \bar{C}} +
2 \partial_\mu \frac{\delta S}{\delta A_\mu^*} = - 2 \partial_\mu \psi^\mu,
\qquad
\frac{\delta S}{\delta \bar{\phi}} +
2 \partial_\mu \frac{\delta S}{\delta \psi_\mu^*} = 0,
\end{equation}
where the terms on the right--hand side, being linear in the quantum fields, 
are classical breakings, i.e., they are not subjected to any specific 
renormalization.

As already emphasized in Sect. 3 the essential reason for imposing the 
non--covariant Landau type gauge relies on the fact that in such a case the 
action $S$ exhibits much larger symmetries than in a covariant 
one. In particular, it can be easily verified that in this gauge the 
dependence of $S$ on the whole set of matter fields is completely fixed by 
the following linearly broken Ward identities 
(see Eqs.~(\ref{3.9}) and (\ref{4.12})),
\begin{align}
\label{4.20}
\frac{\delta S}{\delta \chi^{\dot{A}}} &=
- i (\sigma^\mu)_{\dot{A} B} \partial_\mu \zeta^B - 
2 \bar{\chi}_{\dot{A}} - C \chi_{\dot{A}}^*,
\nonumber
\\
\frac{\delta S}{\delta \bar{\alpha}^{\dot{A}}} &=
i (\sigma^\mu)_{\dot{A} B} \partial_\mu \Bigr(
\frac{\delta S}{\delta \zeta_B^*} + \alpha^B \Bigr) + 
2 \Bigr(
\frac{\delta S}{\delta \bar{\chi}^{\dot{A} *}} + m \bar{\beta}_{\dot{A}} \Bigr) + 
C \Bigr(
\frac{\delta S}{\delta \chi^{\dot{A}}} - \bar{\alpha}_{\dot{A}}^* \Bigr) +
\Bigr(
\frac{\delta S}{\delta C^*} + \phi \Bigr) \chi_{\dot{A}}^*,
\nonumber
\\
\frac{\delta S}{\delta \bar{\chi}_{\dot{A}}} &=
- i (\sigma^\mu)^{\dot{A} B} \partial_\mu \bar{\zeta}_B + 
2 \chi^{\dot{A}} - \bar{\chi}^{\dot{A} *} C,
\nonumber
\\
\frac{\delta S}{\delta \bar{\beta}_{\dot{A}}} &=
i (\sigma^\mu)^{\dot{A} B} \partial_\mu \Bigr(
\frac{\delta S}{\delta \bar{\zeta}^{B *}} + \beta_B \Bigr) - 
2 \Bigr(
\frac{\delta S}{\delta \chi_{\dot{A}}^*} - 
m \bar{\alpha}^{\dot{A}} \Bigr) - \Bigr(
\frac{\delta S}{\delta \bar{\chi}_{\dot{A}}} - 
\bar{\beta}^{\dot{A} *} \Bigr) C - \bar{\chi}^{\dot{A} *} \Bigr(
\frac{\delta S}{\delta C^*} + \phi \Bigr),
\nonumber
\\
\frac{\delta S}{\delta \alpha^A} &=
- i (\sigma^\mu)_{A \dot{B}} \partial_\mu \bar{\alpha}^{\dot{B}} - 
2 \bar{m} \beta_A - \alpha_A^* C,
\nonumber
\\
\frac{\delta S}{\delta \bar{\zeta}^A} &=
- i (\sigma^\mu)_{A \dot{B}} \partial_\mu \Bigr(
\frac{\delta S}{\delta \bar{\alpha}_{\dot{B}}^*} + 
\bar{\chi}^{\dot{B}} \Bigr) + 
2 \bar{m} \Bigr(
\frac{\delta S}{\delta \beta^{A *}} + m \bar{\zeta}_A \Bigr) - \Bigr(
\frac{\delta S}{\delta \alpha^A} + \zeta_A^* \Bigr) C - 
\alpha_A^* \Bigr(
\frac{\delta S}{\delta C^*} + \phi \Bigr),
\nonumber
\\
\frac{\delta S}{\delta \beta_A} &=
- i (\sigma^\mu)^{A \dot{B}} \partial_\mu \bar{\beta}_{\dot{B}} - 
2 \bar{m} \alpha^A + C \beta^{A *},
\nonumber
\\
\frac{\delta S}{\delta \bar{\zeta}_A} &=
- i (\sigma^\mu)^{A \dot{B}} \partial_\mu \Bigr(
\frac{\delta S}{\delta \bar{\beta}^{\dot{B} *}} + \chi_{\dot{B}} \Bigr) + 
2 \bar{m} \Bigr(
\frac{\delta S}{\delta \alpha_A^*} - m \zeta^A \Bigr) - 
C \Bigr(
\frac{\delta S}{\delta \beta_A} + \bar{\zeta}^{A *} \Bigr) + 
\Bigr(
\frac{\delta S}{\delta C^*} + \phi \Bigr) \beta^{A *}.
\end{align}
The stability constraints (\ref{4.18}) -- (\ref{4.20}), which drastically 
reduce the number of independent invariant counterterms, can be established 
to all orders of perturbation theory by using the renormalized quantum action 
principles \cite{18a}.

Moreover, there exists a set of {\rm global} constraints, usually 
valid in the 
Landau type gauge. The first one is the so--called ghost equation \cite{29}, 
which in the present case reads 
\begin{align}
\label{4.21}
\mathbf{G} S = \int& d^4x\, \Bigr\{  
[ A_\mu^*, A^\mu ] + [ \psi_\mu^*, \psi^\mu ] -
[ C^*, C ] + [ \phi^*, \phi ] - T^i  \Bigr( 
\zeta_A^* T^i \zeta^A + \alpha_A^* T^i \alpha^A
\nonumber
\\
& - \bar{\beta}^{\dot{A} *} T^i \bar{\beta}_{\dot{A}} +
\bar{\chi}^{\dot{A} *} T^i \bar{\chi}_{\dot{A}} - 
\bar{\zeta}^{A *} \bar{\zeta}_A T^i - \beta^{A *} \beta_A T^i +
\bar{\alpha}_{\dot{A}}^* \bar{\alpha}^{\dot{A}} T^i -
\chi_{\dot{A}}^* \chi^{\dot{A}} T^i \Bigr) \Bigr\},
\end{align}
with
\begin{equation*}
\mathbf{G} = \int d^4x\, \Bigr\{ 
\frac{\delta }{\delta C} + 
\Bigr[ \bar{C}, \frac{\delta }{\delta B} \Bigr] +
\Bigr[ \bar{\phi}, \frac{\delta }{\delta \eta} \Bigr] + \frac{1}{2} 
\Bigr[ \chi^{\mu\nu}, \frac{\delta }{\delta \lambda^{\mu\nu}} \Bigr]
\Bigr\},
\end{equation*}
where the terms on the right--hand side of that equation are linear
classical breakings, i.e., they will not get radiative 
corrections.

As usual, commuting the ghost equation (\ref{4.21}) with the ST identity
(\ref{4.14}) one gets a further global constraint fulfilled by the action 
$S$, namely the Ward identity for the rigid gauge invariance,
\begin{equation}
\label{4.22}
\mathbf{R} S = 0,
\end{equation}
where $\mathbf{R} = \{ \mathbf{S}_T, \mathbf{G} \}$ denotes the Ward operator 
for rigid gauge transformations in the space of fields and antifields, 
expressing the fact that all the (anti)fields belong to either the adjoint 
or the $\cal R$--representation of the gauge group $G$.

It is known \cite{18a} that the structure of the invariant counterterms entirely
will be governed by a set of classical stability constraints, provided 
they can be extended to all orders of perturbation theory. The proof, that 
the constraints (\ref{4.21}) and (\ref{4.22}) are extendable to any order of 
perturbation theory, can be appreciably simplified by adopting again 
the strategy 
of Ref.~\cite{31}. One associates to each operator $\mathbf{G}$ and 
$\mathbf{R}$ a global ghost, $\gamma$ and $\tau$, respectively, which
take their values in $Lie(G)$, and introduces the 
following operator:
\begin{equation}
\label{4.23}
\mathbf{O}_T = \mathbf{S}_T + {\rm tr} \Bigr\{
\gamma \mathbf{G} + \tau \mathbf{R} +
[ \tau, \gamma ] \frac{\partial}{\partial \gamma} -
\gamma \frac{\partial}{\partial \tau} +
\tau^2 \frac{\partial}{\partial \tau} \Bigr\}.
\end{equation}
One easily verifies that $\mathbf{O}_T$ is nilpotent,
\begin{equation*}
\{ \mathbf{O}_T, \mathbf{O}_T \} = 0.
\end{equation*}
Furthermore, if it can be proven that the integrated cohomology of 
$\mathbf{O}_T$ turns out to be empty, then the integrated cohomology of 
$\mathbf{S}_T$ is empty as well and, besides the ST identity (\ref{4.14}), 
the constraints (\ref{4.20}) and (\ref{4.21}) can be employed to 
single out invariant counterterms. (The proof, that the integrated 
cohomology of $\mathbf{O}_T$ is indeed empty, will be given in Sect. 5.)
The properties of these additional global ghosts are displayzed in Table 7.
\bigskip
\begin{equation*}
\hbox{
\setlength{\extrarowheight}{3pt}
\begin{tabular}{|c|c|c|}
\hline
& $\gamma$ & $\tau$  
\\ 
\hline
ghost number & 2 & 1  
\\ 
\hline
mass dimension & 0 & 0 
\\ 
\hline
parity & even & odd
\\
\hline
\end{tabular}}
\end{equation*}
\bigskip


\bigskip\medskip
\begin{flushleft}
{\large{\bf 5. BRST cohomology: Anomalies and invariant counterterms}}
\end{flushleft}
\bigskip
Let us discuss now the renormalizability of topological QCD 
in the framework of the 
algebraic BRST technique \cite{18a} which allows for a systematic study of the 
quantum extension of the BRST symmetry. In that framework the proof of 
renormalizability is related to the characterization of some cohomology 
classes of the linearized ST--operator $\mathbf{S}_T$, Eq.~(\ref{4.16}), 
which turns out to 
be essential for the (possible absence of) anomalies and the construction 
of the invariant counterterms. Let us recall that both the anomalies 
$\Delta_{\rm A}$ and the invariant counterterms $\Delta_{\rm C}$
of the (quantum) action $S$
are integrated local polynomials in the (anti)fields with (mass) dimension 
four and ghost number, respectively, one and zero.
In addition, they are constrained by the following consistency conditions  
\begin{gather}
\label{5.1}
\mathbf{S}_T \Delta_{\rm A} = 0,
\qquad
\Delta_{\rm A} \neq \mathbf{S}_T \hat{\Delta}_{\rm A},
\qquad
{\rm gh}(\Delta_{\rm A}) = 1,
\\
\intertext{and}
\label{5.2}
\mathbf{S}_Q \Delta_{\rm C} = 0,
\qquad
\frac{\partial \Delta_{\rm C}}{\partial \rho^\mu} = 0,
\quad
\frac{\partial \Delta_{\rm C}}{\partial \xi^\mu} = 0,
\quad
\frac{\partial \Delta_{\rm C}}{\partial \bar{\xi}} = 0, 
\quad
\frac{\partial \Delta_{\rm C}}{\partial \xi} = 0, 
\qquad
{\rm gh}(\Delta_{\rm C}) = 0.
\end{gather}
Therefore, the cohomological relevant solutions of Eqs.~(\ref{5.1}) and 
(\ref{5.2}) are the non--trivial cocycles of the integrated cohomology of 
$\mathbf{S}_T$ and $\mathbf{S}_Q$, respectively. Let us mention that both, 
$\Delta_{\rm A}$ and $\Delta_{\rm C}$, by virtue of 
$[ \mathbf{Z}, \mathbf{S}_T ] = 0$, must have vanishing central charge, i.e., 
$\mathbf{Z} \Delta_{\rm A} = 0$ and $\mathbf{Z} \Delta_{\rm C} = 0$;
analogously for $\mathbf{\bar Z}$.

In order to characterize the integrated cohomology of $\mathbf{S}_T$ we 
introduce the filtration
\begin{equation*}
\mathbf{F}_T = 2 \rho^\mu \frac{\partial}{\partial \rho^\mu} +
\xi^\mu \frac{\partial}{\partial \xi^\mu} - 
\bar{\xi} \frac{\partial}{\partial \bar{\xi}} +
\xi \frac{\partial}{\partial \xi} +
\bar{m} \frac{\partial}{\partial \bar{m}} + 
m \frac{\partial}{\partial m},
\end{equation*}
which obviously, by virtue of $[ m \partial/\partial m, \mathbf{Z} ] =
\mathbf{Z}$, induces a separation of $\mathbf{S}_T$ according to
\begin{equation*}
\mathbf{S}_T = \sum_{n = 0} \mathbf{S}_T^{(n)},
\qquad
[ \mathbf{F}_T, \mathbf{S}_T^{(n)} ] = n \mathbf{S}_T^{(n)},
\end{equation*} 
$\mathbf{S}_T^{(0)} \equiv \mathbf{S}_Q^{(m = 0)}$ being the $m$-- (and 
$\bar{m}$--) independent part of $\mathbf{S}_Q$, Eq.~(\ref{4.x}).
Since for $m = 0$ and $\bar{m} = 0$ the central charges 
vanish the operator $\mathbf{S}_T^{(0)}$ is strictly nilpotent,
\begin{equation*}
\{ \mathbf{S}_T^{(0)}, \mathbf{S}_T^{(0)} \} = 0.
\end{equation*} 
According to the general results on cohomology \cite{33}, the integrated 
cohomology of $\mathbf{S}_T$ is isomorphic to a subspace of the integrated 
cohomology of $\mathbf{S}_T^{(0)}$.

In order to characterize the integrated cohomology of $\mathbf{S}_T^{(0)}$
we introduce the filtration
\begin{align*}
\mathbf{F}_Q = \int d^4x\, {\rm tr} \Bigr\{&
A^\mu \frac{\delta }{\delta A^\mu} +
\psi^\mu \frac{\delta }{\delta \psi^\mu} + 
\frac{1}{2} \chi^{\mu\nu} \frac{\delta }{\delta \chi^{\mu\nu}} +
\frac{1}{2} \lambda^{\mu\nu} \frac{\delta }{\delta \lambda^{\mu\nu}}
\\
& + \bar{C} \frac{\delta }{\delta \bar{C}} +
B \frac{\delta }{\delta B} +
\bar{\phi} \frac{\delta }{\delta \bar{\phi}} +
\eta \frac{\delta }{\delta \eta} + 
2 C \frac{\delta }{\delta C} +
2 \phi \frac{\delta }{\delta \phi} \Bigr\}
\\
+ \int d^4x\, \Bigr\{&
\alpha^A \frac{\delta }{\delta \alpha^A} + 
\bar{\beta}_{\dot{A}} \frac{\delta }{\delta \bar{\beta}_{\dot{A}}} +
\bar{\zeta}^A \frac{\delta }{\delta \bar{\zeta}^A} + 
\chi_{\dot{A}} \frac{\delta }{\delta \chi_{\dot{A}}}
\\
& + \beta_A \frac{\delta }{\delta \beta_A} + 
\bar{\alpha}^{\dot{A}} \frac{\delta }{\delta \bar{\alpha}^{\dot{A}}} +
\zeta_A \frac{\delta }{\delta \zeta_A} +
\bar{\chi}^{\dot{A}} \frac{\delta }{\delta \bar{\chi}^{\dot{A}}} \Bigr\},
\end{align*}
which has the structure of a counting operator. Therefore, $\mathbf{F}_Q$ 
possesses the property of decomposing $\mathbf{S}_T^{(0)}$ as follows:
\begin{equation*}
\mathbf{S}_T^{(0)} = \sum_{n = 0} \mathbf{S}_T^{(n,0)},
\qquad
[ \mathbf{F}_Q, \mathbf{S}_T^{(n,0)} ] = n \mathbf{S}_T^{(n,0)},
\end{equation*} 
where $\mathbf{S}_T^{(0,0)}$ is just that linear, $m$-- and 
$\bar{m}$--independent part of the operator $\mathbf{S}_Q$ which also
does not depend on $S$:
\begin{align*}
\mathbf{S}_T^{(0,0)} = \int d^4x\, {\rm tr} \Bigr\{&
\psi^\mu \frac{\delta }{\delta A^\mu} +
A_\mu^* \frac{\delta }{\delta \psi_\mu^*} +
\phi \frac{\delta }{\delta C} +
C^* \frac{\delta }{\delta \phi^*} 
+ B \frac{\delta }{\delta \bar{C}} +
\eta \frac{\delta }{\delta \bar{\phi}} +
\frac{1}{2} \lambda^{\mu\nu} \frac{\delta }{\delta \chi^{\mu\nu}} 
\Bigr\}
\\
+ \int d^4x\, \Bigr\{&
\alpha^A \frac{\delta }{\delta \bar{\zeta}^A} +
\zeta_A^* \frac{\delta }{\delta \alpha_A^*}  +
\beta_A \frac{\delta }{\delta \zeta_A} +
\bar{\zeta}^{A *} \frac{\delta }{\delta \beta^{A *}} 
\\
& + \chi_{\dot{A}} \frac{\delta }{\delta \bar{\beta}_{\dot{A}}} +
\bar{\beta}^{\dot{A} *} \frac{\delta }{\delta \bar{\chi}^{\dot{A} *}} +
\bar{\chi}^{\dot{A}} \frac{\delta }{\delta \bar{\alpha}^{\dot{A}}} +
\bar{\alpha}_{\dot{A}}^* \frac{\delta }{\delta \chi_{\dot{A}}^*} 
\Bigr\}.
\end{align*}

Since in $\mathbf{S}_T^{(0,0)}$ all the (anti)fields appear in 
BRST--doublets, one concludes that the integrated cohomology of
$\mathbf{S}_T^{(0,0)}$ is empty \cite{32,33}. Thus, the integrated cohomology 
of the operator $\mathbf{S}_T$ is empty as well, due to the fact that it is, 
in turn, isomorphic to a subspace of the integrated cohomology of 
$\mathbf{S}_T^{(0,0)}$. This result implies that the general solutions of the 
consistency conditions (\ref{5.1}) and (\ref{5.2}) are given by
\begin{gather}
\Delta_{\rm A} = 0,
\nonumber\\
\intertext{and}
\label{5.3}
\Delta_{\rm C} = \mathbf{S}_Q \hat{\Delta}_{\rm C},
\qquad
{\rm gh}(\hat{\Delta}_{\rm C}) = -1, 
\end{gather}
where $\hat{\Delta}_{\rm C}$ is the most general integrated local polynomial
 in the (anti)fields, with dimension $7/2$ and ghost number minus one.  
Hence, the ST identity (\ref{4.14}) is anomaly free and the invariant 
counterterms are trivial cocycles of the integrated cohomology of 
$\mathbf{S}_Q$. 

The absence of anomalies of the ST identity,
$\mathbf{S}_Q(S) = 0$, as well as of the Ward identities 
$\mathbf{Q}_\mu S = 0$, $\mathbf{P}_\mu S = 0$, $\mathbf{Z} = 0$ and 
$\mathbf{\bar{Z}} S = 0$,  with $S$ 
being independent of $\rho^\mu$, $\xi^\mu$, $\bar{\xi}$ and $\xi$, implies 
that $\Delta_{\rm C}$, besides (\ref{5.2}), is subjected also to the following 
consistency conditions,
\begin{equation}
\label{5.4}
\mathbf{Q}_\mu \Delta_{\rm C} = 0,
\qquad
\mathbf{P}_\mu \Delta_{\rm C} = 0,
\qquad
\mathbf{Z} \Delta_{\rm C} = 0,
\qquad
\mathbf{\bar{Z}} \Delta_{\rm C} = 0.
\end{equation}
Furthermore, from the ghost equation (\ref{4.21}) and the Ward identity
(\ref{4.22}), which can be imposed to any order of perturbation theory, 
it follows that $\Delta_{\rm C}$ is required to obey the following
constraints, too,
\begin{equation}
\label{5.5}
\mathbf{G} \Delta_{\rm C} = 0,
\qquad
\mathbf{R} \Delta_{\rm C} = 0.
\end{equation}

In order to prove that the stability constraints (\ref{4.21}) and (\ref{4.22}) 
can be extended to the quantum level it is sufficient to verify that the 
integrated cohomology of the nilpotent operator $\mathbf{O}_T$, 
Eq.~(\ref{4.23}), is empty. For that purpose, let us introduce the filtration
\begin{equation*}
\mathbf{N}_T = \mathbf{F}_T + 
{\rm tr} \Bigr\{ 2 \gamma \frac{\partial}{\partial \gamma} +
2 \tau \frac{\partial}{\partial \tau} \Bigr\} 
\end{equation*}
which clearly induces a separation of $\mathbf{O}_T$, namely
\begin{gather}
\mathbf{O}_T = \sum_{n = 0} \mathbf{O}_T^{(n)},
\qquad
[ \mathbf{N}_T, \mathbf{O}_T^{(n)} ] = n \mathbf{O}_T^{(n)},
\end{gather}
\vspace*{-.5cm}
{with}
\begin{gather}
\label{5.7}
\mathbf{O}_T^{(0)} = \mathbf{S}_T^{(0)} -
{\rm tr} \Bigr\{ \gamma \frac{\partial}{\partial \tau} \Bigr\}.
\end{gather}
Since the integrated cohomology of $\mathbf{S}_Q(0)$ is empty and because
the extra term in (\ref{5.7}) appears as BRST--doublet the integrated
cohomolgy of $\mathbf{O}_T^{(0)}$ is empty as well and, therefore, 
with the same reasoning as before, the integrated cohomology of $\mathbf{O}_T$ 
is empty, too. 

From the stability constraints (\ref{4.18}) -- (\ref{4.20}) for $\Delta_{\rm C}$ 
we get a further set of restrictions:  
\begin{equation}
\label{5.8}
\frac{\delta \Delta_{\rm C}}{\delta B} = 0,
\qquad
\frac{\delta \Delta_{\rm C}}{\delta \bar{C}} + 
2 \partial_\mu \frac{\delta \Delta_{\rm C}}{\delta A_\mu^*} = 0,
\qquad
\frac{\delta \Delta_{\rm C}}{\delta \eta} = 0,
\qquad
\frac{\delta \Delta_{\rm C}}{\delta \bar{\phi}} + 
2 \partial_\mu \frac{\delta \Delta_{\rm C}}{\delta \psi_\mu^*} = 0
\end{equation}
and
\begin{align}
\label{5.9}
\frac{\delta \Delta_{\rm C}}{\delta \chi^{\dot{A}}} &= 0,
\qquad
\frac{\delta \Delta_{\rm C}}{\delta \bar{\alpha}^{\dot{A}}} -
i (\sigma^\mu)_{\dot{A} B} \partial_\mu 
\frac{\delta \Delta_{\rm C}}{\delta \zeta_B^*} - 
2 \frac{\delta \Delta_{\rm C}}{\delta \bar{\chi}^{\dot{A} *}} -
C \frac{\delta \Delta_{\rm C}}{\delta \chi^{\dot{A}}} -
\frac{\delta \Delta_{\rm C}}{\delta C^*} \chi_{\dot{A}}^* = 0,
\nonumber
\\
\frac{\delta \Delta_{\rm C}}{\delta \bar{\chi}_{\dot{A}}} &= 0,
\qquad
\frac{\delta \Delta_{\rm C}}{\delta \bar{\beta}_{\dot{A}}} -
i (\sigma^\mu)^{\dot{A} B} \partial_\mu 
\frac{\delta \Delta_{\rm C}}{\delta \bar{\zeta}^{B *}} + 
2 \frac{\delta \Delta_{\rm C}}{\delta \chi_{\dot{A}}^*} +
\frac{\delta \Delta_{\rm C}}{\delta \bar{\chi}_{\dot{A}}} C +
\bar{\chi}^{\dot{A} *} \frac{\delta \Delta_{\rm C}}{\delta C^*} = 0,
\nonumber
\\
\frac{\delta \Delta_{\rm C}}{\delta \alpha^A} &= 0,
\qquad
\frac{\delta \Delta_{\rm C}}{\delta \zeta^A} +
i (\sigma^\mu)_{A \dot{B}} \partial_\mu 
\frac{\delta \Delta_{\rm C}}{\delta \bar{\alpha}_{\dot{B}}^*} - 
2 \bar{m} \frac{\delta \Delta_{\rm C}}{\delta \beta^{A *}} +
\frac{\delta \Delta_{\rm C}}{\delta \alpha^A} C +
\alpha_A^* \frac{\delta \Delta_{\rm C}}{\delta C^*} = 0,
\nonumber
\\
\frac{\delta \Delta_{\rm C}}{\delta \beta_A} &= 0,
\qquad
\frac{\delta \Delta_{\rm C}}{\delta \bar{\zeta}_A} +
i (\sigma^\mu)^{A \dot{B}} \partial_\mu 
\frac{\delta \Delta_{\rm C}}{\delta \bar{\beta}^{\dot{B} *}} - 
2 \bar{m} \frac{\delta \Delta_{\rm C}}{\delta \alpha_A^*} +
C \frac{\delta \Delta_{\rm C}}{\delta \beta^A} - 
\frac{\delta \Delta_{\rm C}}{\delta C^*} \beta^{A *} = 0.
\end{align}

The constraints (\ref{5.4}) imply that $\Delta_{\rm C}$ is required to be 
invariant under vector supersymmetry transformations, space--time 
translations and central charge transformations. From (\ref{5.5}) one infers 
that the ghost $C$ enters into $\Delta_{\rm C}$ either as derivative 
$\partial_\mu C$ or through the combinations 
\begin{align*}
B - \{ C, \bar{C} \}, 
\qquad
\eta - [ C, \bar{\phi} ] 
\qquad 
\hbox{and}
\qquad
\lambda^{\mu\nu} - \{ C, \chi^{\mu\nu} \},
\end{align*}
 respectively, and that 
$\Delta_{\rm C}$ can be taken to be rigid gauge invariant.

Concerning the constraints (\ref{5.8}) it follows 
that the auxiliary fields $B$ and $\eta$ cannot appear in $\Delta_{\rm C}$ 
and that the antighosts $\bar{C}$ and $\bar{\phi}$ can enter only through 
the combinations 
\begin{align*}
A_\mu^* - 2 \partial_\mu \bar{C}
\qquad 
\hbox{and} 
\qquad 
\psi_\mu^* - 2 \partial_\mu \bar{\phi},
\end{align*}
respectively. Furthermore, 
from (\ref{5.9}) it follows that the matter fields $\bar{\chi}^{\dot{A}}$, 
$\chi_{\dot{A}}$, $\alpha^A$ and $\beta_A$ cannot appear in 
$\Delta_{\rm C}$ and that
$\bar{\alpha}^{\dot{A}}$, $\bar{\beta}_{\dot{A}}$, $\bar{\zeta}^A$ and
$\zeta_A$ can enter only through the following combinations: 
\begin{alignat*}{3}
& \zeta_A^* - i (\sigma^\mu)_{A \dot{B}} 
\partial_\mu \bar{\alpha}^{\dot{B}},
&\qquad&
\bar{\chi}^{\dot{A}*} + 2 \bar{\alpha}^{\dot{A}},
&\qquad&
C^* + \bar{\chi}_{\dot{A}} \bar{\alpha}^{\dot{A}}, 
\\
& \bar{\zeta}^{A *} - i (\sigma^\mu)^{A \dot{B}} 
\partial_\mu \bar{\beta}_{\dot{B}},
&\qquad&
\chi_{\dot{A}}^* - 2 \beta_{\dot{A}},
&\qquad&
C^* - \bar{\beta}_{\dot{A}} \bar{\chi}^{\dot{A} *},
\\ 
& \bar{\alpha}_{\dot{A}}^* + i (\sigma^\mu)_{\dot{A} B} 
\partial_\mu \zeta^B,
&\qquad&
\beta^{A *} + 2 \bar{m} \zeta^A,
&\qquad&
C^* - \zeta^A \alpha_A^*,
\\ 
& \bar{\beta}^{\dot{A} *} + i (\sigma^\mu)^{\dot{A} B} 
\partial_\mu \bar{\zeta}_B,
&\qquad&
\alpha_A^* + 2 \bar{m} \bar{\zeta}_A, 
&\qquad&
C^* - \beta^{A *} \bar{\zeta}_A.
\end{alignat*}

Let us now turn to the computation of $\hat{\Delta}_{\rm C}$. First of all
we point out that the whole set of constraints 
is stable under the action of $\mathbf{S}_Q$ and, therefore, 
$\Delta_{\rm C} = \mathbf{S}_Q \hat{\Delta}_{\rm C}$ satisfies these 
constraints if $\hat{\Delta}_{\rm C}$ will do. 
Although, generally, $\hat{\Delta}_{\rm C}$ 
does not have to obey them it may be chosen to do so, except for the 
constraint $\mathbf{G} \Delta_{\rm C} = 0$ \cite{17}. Since the constraints 
above are very restrictive they will give rise to a rather small set of 
independent counterterms. Indeed, the only  possible invariant 
counterterms which are combatible with the constraints 
(\ref{5.4}), (\ref{5.5}) and (\ref{5.8}) are obtained from 
$\Delta_{\rm C} = \mathbf{S}_Q \hat{\Delta}_{\rm C}$ by choosing for 
$\hat{\Delta}_{\rm C}$ the following seven combinations of fields 
and antifields, 
\begin{align*}
\hat{\Delta}_{\rm C} = \int d^4x\, {\rm tr} \Bigr\{&
z_1 (A_\mu^* - 2 \partial_\mu \bar{C} ) A^\mu +
z_1 (\psi_\mu^* - 2 \partial_\mu \bar{\phi} ) \psi^\mu
\\
& +  z_2 (\psi_\mu^* - 2 \partial_\mu \bar{\phi} ) \partial^\mu C + 
z_3 \chi^{\mu\nu} \partial_\mu A_\nu +
z_4 \chi^{\mu\nu} [ A_\mu, A_\nu ] \Bigr\}
\\ 
+ \int d^4x\, \Bigr\{&
( z_5 \bar{\chi}^{\dot{A} *} + z_7 2 \bar{\alpha}^{\dot{A}} ) 
\bar{\chi}_{\dot{A}} - ( z_5 \bar{\beta}^{\dot{A} *} + 
z_7 i (\sigma^\mu)^{\dot{A} B} \partial_\mu \bar{\zeta}_B ) 
\bar{\beta}_{\dot{A}}
\\
& + ( z_6 \chi_{\dot{A}}^* - z_7 2 \beta_{\dot{A}} ) 
\chi^{\dot{A}} - ( z_6 \bar{\alpha}_{\dot{A}}^* +
z_7 i (\sigma^\mu)_{\dot{A} B} \partial_\mu \zeta^B ) 
\bar{\alpha}^{\dot{A}}
\\
& + ( z_5 \alpha_A^* + z_7 2 \bar{m} \bar{\zeta}_A ) \alpha^A + 
( z_5 \zeta_A^* - 
z_7 i (\sigma^\mu)_{A \dot{B}} \partial_\mu \bar{\alpha}^{\dot{B}} ) 
\zeta^A
\\
& + ( z_6 \beta^{A *} + z_7 2 \bar{m} \zeta^A ) \beta_A + 
( z_6 \bar{\zeta}^{A *} -
z_7 i (\sigma^\mu)^{A \dot{B}} \partial_\mu \bar{\beta}_{\dot{B}} )
\bar{\zeta}_A \Bigr\},
\end{align*}
where $z_1, z_2, \cdots, z_7$ are arbitrary coefficients. This set of 
independent counterterms is reduced further by applying the last constraints 
(\ref{5.9}), leading to some relations among the renormalization factors, 
namely $z_5 = z_7$, $z_6 = z_7$ and $z_7 = 0$, respectively. Hence, one ends 
up with only four possible invariant counterterms,
\begin{align}
\label{5.10}
\hat{\Delta}_{\rm C} = \int d^4x\, {\rm tr} \Bigr\{&
z_1 ( A_\mu^* - 2 \partial_\mu \bar{C} ) A^\mu +
z_1 ( \psi_\mu^* - 2 \partial_\mu \bar{\phi} ) \psi^\mu
\nonumber
\\
& + z_2 ( \psi_\mu^* - 2 \partial_\mu \bar{\phi} ) \partial^\mu C + 
z_3 \chi^{\mu\nu} \partial_\mu A_\nu +
z_4 \chi^{\mu\nu} [ A_\mu, A_\nu ] \Bigr\}.
\end{align}

Here, some remarks are in order. First, since all the quantum corrections 
to the action $S$ are trivial $\mathbf{S}_Q$--cocycles there appears no 
physical coupling parameter in TQCD and therefore, the coefficients 
$z_1, z_2, z_3$ and $z_4$ are anomalous dimensions 
redefinig the fields and antifields. Second, an unexpected feature 
of the non--covariant Landau type gauge adopted 
here is the fact that the independent counterterms of TQCD 
agree with those of TYM. 
This means, on the one hand, that the matter action of TQCD is finite, 
i.e., it does not receive any radiative corrections from higher loops, and, 
on the other hand, that the requirement of the ghost for the ghost 
equation is not actually necessary. Third, as already pointed out in 
Ref.~\cite{20,17}, as a consequence of the fact that in the Landau type 
gauge, $\xi = 0$, is stable under renormalization 
(i.e., that only in this gauge $\xi$ receives no renormalization),
the counterterm
\begin{equation*}
\Delta_{\rm C} \sim \int d^4x\, {\rm tr} \Bigr\{ 
F_{\mu\nu}^+ F^{\mu\nu +} \Bigr\},
\end{equation*}
disappears. This result is essential in preserving the topological nature of
the model. Furthermore, the absence of that counterterm guarantees that 
the $\beta$--function vanishes. This is in accordance with an one--loop 
computation carried out in Refs.~\cite{7,1}.

\newpage

\bigskip\medskip
\begin{flushleft}
{\large{\bf 6. Concluding remarks}}
\end{flushleft}
\bigskip
In this paper we have studied the renormalizability of twisted $N = 2$ 
supersymmetry with $Z = 2$ central charges. By coupling the gauge multiplet
to the standard massive hypermultiplet, i.e., with only one central charge,
the $R$--symmetry is broken into $Z_2$. In such a case we are faced with the 
situation that the ghost number of the gauge--fixed action and, in 
consequence of this, the cohomology classes of the counter terms and 
anomalies are not 
uniquely characterized. Here, it has been shown that this problem can be
avoided by introducing two central charges $Z$ and $\bar{Z}$, being 
complex conjugate to each other, and formally ascribing to them, 
as well as to their
eigenvalues $\pm m$ and $\pm \bar{m}$, the $R$--weights (ghost numbers) 
$R(Z) = R(m) = 2 R(Q)$ and $R(\bar{Z}) = R(\bar{m}) = 2 R(\bar{Q}_\mu)$.

By adopting the non--covariant Landau type gauge and by making use of both 
the topological shift symmetry $Q$ and the vector supersymmetry $\bar{Q}$ 
it has been proven that the twisted hypermultiplets are not subjected to any 
renormalization, i.e., the matter action of TQCD is perturbatively finite. 
Thus, in that particular gauge, which should be as acceptable as any other
gauge choice, TQCD is renormalizable with the 
same counter terms as TYM. Since the invariant counterterms are trivial 
BRST--cocycles no physical parameters appear in that model.

In this paper we have not analyzed the question whether the choice of a 
non--covariant gauge, which significantly differs from the original one of 
TYM, may eventually change the construction of the topological observables of 
that model.

Furthermore, no attention has been paid to a possible nontrivial
$\theta$ term in the 
topological action. The question whether this term, without loss of 
generality, can be droped at the beginning or whether the $\theta$ angle 
does receive radiation corrections from higher loops will be studied
elsewhere.


\bigskip\medskip
\begin{flushleft}
{\large{\bf Appendix A: Two--spinor notations in Euclidean space--time}}
\end{flushleft}
\bigskip
The Weyl 2--spinor conventions in Euclidean space--time
adopted in this paper are those of App.~E in Ref.~\cite{34}. 
The matrices $(\sigma_\mu)^{A \dot{B}}$ and $(\sigma_\mu)_{\dot{A} B}$, 
being invariant numerical tensors of $SL(2,C)$ if 
$\mu$ transforms according to the vector representation
of $SO(4)$, are defined by   
\begin{equation*}
(\sigma_\mu)^{A \dot{B}} = ( \sigma_\alpha, i {\rm I}_2 )^{A \dot{B}}, 
\qquad
(\sigma_\mu)_{\dot{A} B} = ( \sigma_\alpha, - i {\rm I}_2 )_{\dot{A} B},
\end{equation*}
with $\sigma_\alpha$ ($\alpha = 1,2,3$) being the 
Pauli matrices.

Another set of invariant tensors are the antisymmetric matrices
$\epsilon_{AB}$ and $\epsilon_{\dot{A}\dot{B}}$,
\begin{equation*}
\epsilon_{AB} = 
\begin{pmatrix} 0 & 1 \\ - 1 & 0 \end{pmatrix} = \epsilon^{AB}
\qquad
\epsilon_{\dot{A}\dot{B}} =
\begin{pmatrix} 0 & - 1 \\ 1 & 0 \end{pmatrix} = \epsilon^{\dot{A}\dot{B}},
\end{equation*}
with 
$\epsilon^{AB}$ and $\epsilon^{\dot{A}\dot{B}}$ being defined by 
$\epsilon^{AC} \epsilon_{CB} = - \delta^A_{\!~~B}$ and 
$\epsilon^{\dot{A}\dot{C}} \epsilon_{\dot{C}\dot{B}} = 
- \delta^{\dot{A}}_{\!~~\dot{B}}$. These tensors raise and lower the spinor
indices according to  
$\psi_A = \psi^B \epsilon_{BA}$,
$\psi^A = \epsilon^{AB} \psi_B$ and 
$\psi_{\dot{A}} = \psi^{\dot{B}} \epsilon_{\dot{B}\dot{A}}$,
$\psi^{\dot{A}} = \epsilon^{\dot{A}\dot{B}} \psi_{\dot{B}}$, respectively.

The matrices $(\sigma_\mu)^{A \dot{B}}$ and $(\sigma_\mu)_{\dot{A} B}$
satisfy the Clifford algebra
\begin{align*}
(\sigma_\mu)^{A \dot{C}} (\sigma_\nu)_{\dot{C} B} +
(\sigma_\nu)^{A \dot{C}} (\sigma_\mu)_{\dot{C} B} &= 
2 \delta_{\mu\nu} \delta^A_{\!~~B},
\\
(\sigma_\mu)_{\dot{A} C} (\sigma_\nu)^{C \dot{B}} +
(\sigma_\nu)_{\dot{A} C} (\sigma_\mu)^{C \dot{B}} &= 
2 \delta_{\mu\nu} \delta^{\dot{A}}_{\!~~\dot{B}},
\end{align*}
and, in addition, the completeness relations
\begin{align*}
(\sigma_\mu)_{\dot{A} B} (\sigma^\nu)^{B \dot{A}} &=
2 \delta_\mu^\nu,
\\
(\sigma_\mu)_{\dot{A} B} (\sigma^\mu)_{\dot{C} D} =
2 \epsilon_{\dot{A}\dot{C}} \epsilon_{BD},
\qquad
(\sigma_\mu)_{\dot{A} B} (\sigma^\mu)^{C \dot{D}} &=
2 \delta_{\dot{A}}^{\dot{D}} \delta_B^C,
\qquad
(\sigma_\mu)^{A \dot{B}} (\sigma^\mu)^{C \dot{D}} =
2 \epsilon^{AC} \epsilon^{\dot{B}\dot{D}}.
\end{align*}
Since $(\sigma_\mu)^{A \dot{B}}$ is the hermitean conjugate of
$(\sigma_\mu)_{A \dot{B}}$, it holds
$(\sigma_\mu)^{A \dot{B}} \equiv \epsilon^{AC} \epsilon^{\dot{B}\dot{D}}
(\sigma_\mu)_{\dot{D} C} = (\sigma_\mu)^{\dot{B} A}$ and, lowering its 
indices, $(\sigma_\mu)_{A \dot{B}} = (\sigma_\mu)_{\dot{B} A}$. Hence,
$(\sigma_\mu)^{A \dot{B}}$ and $(\sigma_\mu)_{A \dot{B}}$ are symmetric in
their spinor indices.

The self--dual and antiself--dual $SO(4)$ generators 
$(\sigma_{\mu\nu})^{AB}$ and $(\sigma_{\mu\nu})_{\dot{A}\dot{B}}$, 
respectively, being antisymmetric in their vector
indices and symmetric in their spinor indices, are defined by
\begin{align*}
(\sigma_{\mu\nu})^{AB} &= (\sigma_\mu)^{A \dot{C}} 
(\sigma_\nu)_{\dot{C}}^{\!~~B} + \delta_{\mu\nu} \epsilon^{AB},
\qquad
(\sigma_{\mu\nu})^{AB} = + 
(\tilde{\sigma}_{\mu\nu})^{AB}
\\
(\sigma_{\mu\nu})_{\dot{A}\dot{B}} &= (\sigma_\mu)_{\dot{A} C} 
(\sigma_\nu)^C_{\!~~\dot{B}} - \delta_{\mu\nu} \epsilon_{\dot{A}\dot{B}},
\qquad
(\sigma_{\mu\nu})_{\dot{A}\dot{B}} = - 
(\tilde{\sigma}_{\mu\nu})_{\dot{A}\dot{B}}.
\end{align*}

A vector $V_\mu$ and an antisymmetric tensor $T_{\mu\nu}$ are represented by
\begin{equation*}
V_\mu = - \hbox{\large$\frac{1}{2}$} (\sigma_\mu)^{A \dot{B}} V_{A \dot{B}},
\qquad
V_{\dot{A} B} = (\sigma^\mu)_{\dot{A} B} V_\mu,
\end{equation*}
and
\begin{equation*}
T_{\mu\nu} = \hbox{\large$\frac{1}{4}$} (\sigma_\mu)^{A \dot{B}} (\sigma_\nu)^{C \dot{D}}
T_{A \dot{B} C \dot{D}} = T_{\mu\nu}^+ + T_{\mu\nu}^-,
\qquad
T_{A \dot{B} C \dot{D}} = 
\epsilon_{\dot{B}\dot{D}} T_{AC}^+ +
\epsilon_{AC} T_{\dot{B}\dot{D}}^-,
\end{equation*}
respectively, where $T_{\mu\nu}^\pm=\pm (\sigma_{\mu\nu})^{AB}T_{AB}^\pm$ 
is the (anti)self--dual part of $T_{\mu\nu}$.

Finally, some often used identities are
\begin{align*}
(\sigma_\rho)^{A \dot{C}} (\sigma_{\mu\nu})_{\dot{C}}^{\!~~\dot{B}} &=
\delta_{\rho\mu} (\sigma_\nu)^{A \dot{B}} -
\delta_{\rho\nu} (\sigma_\mu)^{A \dot{B}} -
\epsilon_{\mu\nu\rho\sigma} (\sigma^\sigma)^{A \dot{B}},
\\
(\sigma_\rho)_{\dot{A} C} (\sigma_{\mu\nu})^C_{\!~~B} &=
\delta_{\rho\mu} (\sigma_\nu)_{\dot{A} B} -
\delta_{\rho\nu} (\sigma_\mu)_{\dot{A} B} +
\epsilon_{\mu\nu\rho\sigma} (\sigma^\sigma)_{\dot{A} B},
\\
(\sigma_{\mu\nu})^{\dot{A} \dot{C}} 
(\sigma_{\rho\sigma})_{\dot{C}}^{\!~~\dot{B}} &=
2 ( \delta_{\mu\rho} \delta_{\nu\sigma} -
\delta_{\nu\rho} \delta_{\mu\sigma} - \epsilon_{\mu\nu\rho\sigma} )
\epsilon^{\dot{A} \dot{B}}
\\
&\quad - \delta_{\mu\rho} (\sigma_{\nu\sigma})^{\dot{A} \dot{B}} +
\delta_{\nu\rho} (\sigma_{\mu\sigma})^{\dot{A} \dot{B}} -
\delta_{\nu\sigma} (\sigma_{\mu\rho})^{\dot{A} \dot{B}} +
\delta_{\mu\sigma} (\sigma_{\nu\rho})^{\dot{A} \dot{B}},
\\
(\sigma_{\mu\nu})_{AC} 
(\sigma_{\rho\sigma})^C_{\!~~B} &=
2 ( \delta_{\mu\rho} \delta_{\nu\sigma} -
\delta_{\nu\rho} \delta_{\mu\sigma} + \epsilon_{\mu\nu\rho\sigma} )
\epsilon_{AB}
\\
&\quad - \delta_{\mu\rho} (\sigma_{\nu\sigma})_{AB} +
\delta_{\nu\rho} (\sigma_{\mu\sigma})_{AB} -
\delta_{\nu\sigma} (\sigma_{\mu\rho})_{AB} +
\delta_{\mu\sigma} (\sigma_{\nu\rho})_{AB}
\end{align*}
and
\begin{align*}
(\sigma_{\mu\nu})^{\dot{A}\dot{B}} (\sigma_{\rho\sigma})_{\dot{A}\dot{B}} &=
2 ( \delta_{\mu\rho} \delta_{\nu\sigma} -
\delta_{\nu\rho} \delta_{\mu\sigma} - \epsilon_{\mu\nu\rho\sigma} ),
\qquad\!\!
(\sigma^{\mu\nu})^{\dot{A}\dot{B}} (\sigma_{\mu\nu})^{\dot{C}\dot{D}} =
8 \epsilon^{\dot{A}\dot{C}} \epsilon^{\dot{B}\dot{D}} -
4 \epsilon^{\dot{A}\dot{B}} \epsilon^{\dot{C}\dot{D}},
\\
(\sigma_{\mu\nu})_{AB} (\sigma_{\rho\sigma})^{AB} &=
2 ( \delta_{\mu\rho} \delta_{\nu\sigma} -
\delta_{\nu\rho} \delta_{\mu\sigma} + \epsilon_{\mu\nu\rho\sigma} ),
\qquad\!\!
(\sigma^{\mu\nu})_{AB} (\sigma_{\mu\nu})_{CD} =
8 \epsilon_{AC} \epsilon_{BD} - 4 \epsilon_{AB} \epsilon_{CD};
\end{align*}
here,  the antisymmetric tensor $\epsilon_{\mu\nu\rho\sigma}$
is normalized according to $\epsilon_{1234} = 1$.


\bigskip\medskip
\begin{flushleft}
{\large{\bf APPENDIX B: Twisting of N = 2 supersymmetric theories
with two central charges $Z$ and $\bar Z$}}
\end{flushleft}
\bigskip
In this Appendix we use the method of dimensional reduction in order to 
include central charges in the superalgebra of $N = 2$ SYM coupled 
to two (massive) hypermultiplets (in the fundamental and its conjugate
representation of $SL(2,C)$).

As is well known there exists a close relationship between extended $N = 2$ 
SYM in $D = 4$ dimensions and simple $N = 1$ SYM with gauge multiplet
$(A_M, \lambda), M= 1, \cdots, 6,$ in $D=6$ dimensions. The 
transition from the latter to the former is achieved by a {\it trivial} 
dimensional reduction, namely by demanding that the gauge potential $A_M$ 
 and the chiral spinors $\lambda,\ \bar{\lambda}$ is 
independent of the extra dimensions $x^5$ and $x^6$. After that dimensional   
reduction the extra components of $A_M$ simply become complex scalar 
fields, $A_5 = - i ( \phi + \bar{\phi} )$ and $A_6 = \bar{\phi} - \phi$,
and the rotation group in $D = 6$ dimensional Euclidean space--time 
is broken down according to
$SO(6) \supset SO(4) \otimes SO(2)$. After reduction the chiral fields 
transform under the spinor representation of the universal covering group 
$SL(2,C) \otimes U_R(1)$. (Since $\lambda$ and $\bar\lambda$ are
not subjected to a symplectic 
reality condition the $Sp(2)$ internal symmetry of the $N = 2$ SYM is not 
accounted for in Ref.~\cite{22}. This problem can be circumvented
by reformulating $N=1$ SYM in such manner that the $Sp(2)$ symmetry
will be manifest (see, below).)

There exist also {\it non--trivial} dimensional reductions which allow to
generate also central charges 
in both the massive matter multiplets \cite{35} and the massive ghost
excitations \cite{36}. The central charge $Z$ in the standard massive
hypermultiplet \cite{25} occurs by only compactifying the sixth dimension
$x^6$ into a circle \cite{26} and reducing the fifth dimension trivially.
In order to get two central charges $Z$ and
$\bar{Z}$, being complex conjugate to each other, one has to compactify
the extra dimensions $x^5$ and $x^6$ into a torus and to assume that the 
complex scalar fields $\zeta$, $\bar{\zeta}$ and the anti-chiral Dirac 
spinor $\psi$ of which the (originally massless)
hypermultiplet consists are periodic in $x^5$ and $x^6$ with the (inverse) 
periods $m_5 = m + \bar{m}$ and $m_6 = i(\bar{m} - m)$, 
respectively \cite{26}. Thereby, the central charges are 
identified with the extra components of the generator of space--time 
translations according to $P_5 = Z + \bar{Z}$ and $P_6 = i (\bar{Z} - Z)$, 
respectively.

In order to implement the central charges $Z$ and $\bar{Z}$ into
the matter multiplet let us start from
$N = 1$ SYM coupled to a (massless) hypermultiplet 
in $D = 6$ dimensional Euclidean space--time in Wess-Zumino gauge:
\begin{equation}
\tag{B.1}
W^{(D = 6)} = W_{\rm SYM}^{(N = 1)} + W_{\rm M}^{(Z = 0)},
\end{equation} 
which is built from an antihermitean vector field $A_M$ ($M = 1, \cdots, 6$)
and a $Sp(2)$--doublet of chiral (symplectic) Majorana spinors \cite{37}
$\lambda_a$ ($a = 1,2$) in the adjoint representation of the gauge group,
\begin{equation*}
W_{\rm SYM}^{(N = 1)} = \int d^6x\, {\rm tr} \Bigr\{
\hbox{\large$\frac{1}{4}$} F^{MN} F_{MN} -
\hbox{\large$\frac{1}{2}$} i \bar{\lambda}^a \Gamma^M D_M \lambda_a \Bigr\},
\end{equation*}
and from a $Sp(2)$--doublet of complex scalar fields $\zeta_a$, 
$\bar{\zeta}^a \equiv \zeta_a^\dagger$, and an anti-chiral Dirac spinor 
$\psi$ in some representation $\cal R$ (with generators $T^i$), e.g., 
the fundamental representation, of the gauge group,
\begin{align*}
W_{\rm M}^{(Z = 0)} = \int d^6x\, \Bigr\{ 
i \bar{\psi} \Gamma^M {D_M}  \psi -
i \bar{\psi} \lambda_a \zeta^a + i \bar{\zeta}_a \bar{\lambda}^a \psi - 
\hbox{\large$\frac{1}{2}$}  (\bar{\zeta}_a {D}_M^2 \zeta^a) -
\hbox{\large$\frac{1}{4}$} 
( \bar{\zeta}_a T^i \zeta^b )( \bar{\zeta}_b T^i \zeta^a ) \Bigr\}.
\end{align*} 
with
\begin{align*}
{D}_M = {\partial}_M + A_M^i T^i.
\end{align*}
Here, the 8--dimensional Dirac matrices $\Gamma_M$ and $\Gamma_7$ are 
represented as follows:
\begin{align*}
\Gamma_\mu &= \gamma_\mu \otimes {\rm I}_2,
\qquad
\gamma_\mu = \begin{pmatrix} 0_2 & - (\sigma_\mu)^{A \dot{B}} \\
(\sigma_\mu)_{\dot{A} B} & 0_2 \end{pmatrix},
\qquad
\mu = 1,2,3,4,
\\
\Gamma_{4 + \alpha} &= \gamma_5 \otimes \sigma_\alpha,
\qquad
\gamma_5 = - i \begin{pmatrix} \delta^A_{\!~~B} & 0_2 \\
0_2 & - \delta_{\dot{A}}^{\!~~\dot{B}} \end{pmatrix}, 
\qquad
\alpha = 1,2,3,
\end{align*}
where $\gamma_\mu$ and $\gamma_5$ are the (usual) 4--dimensional Dirac
matrices and $\sigma_\alpha$ are the Pauli matrices. They obey the 
relations
\begin{equation*}
\{ \Gamma_M, \Gamma_N \} = - 2 \delta_{MN} {\rm I}_8,
\qquad
\Sigma_{MN} = - \frac{1}{2} [ \Gamma_M, \Gamma_N ],
\end{equation*}
with
\begin{align*}
\Sigma_{\mu\nu} &= \sigma_{\mu\nu} \otimes {\rm I}_2,
\qquad 
\sigma_{\mu\nu} = \begin{pmatrix} (\sigma_{\mu\nu})^A_{\!~~B} & 0_2 \\
0_2 & (\sigma_{\mu\nu})_{\dot{A}}^{\!~~\dot{B}} \end{pmatrix},
\qquad
\mu, \nu = 1,2,3,4,
\\
\Sigma_{\mu,4 + \alpha} &= - \gamma_\mu \gamma_5 \otimes \sigma_\alpha,
\qquad
\Sigma_{4 + \alpha,4 + \beta} = {\rm I}_4 \otimes 
\hbox{\large$\frac{1}{2}$} [ \sigma_\alpha, \sigma_\beta ],
\qquad
\alpha, \beta = 1,2,3.
\end{align*}

In order to ensure that the action (B.1) is manifestly invariant under the 
internal symmetry group $Sp(2,R)\cong SU(2)$ the chiral 8--spinors $\lambda_a$
and $\bar{\lambda}^a$ are required to obey both the Weyl condition 
$\lambda_a = i \Gamma_7 \lambda_a$ (chirality condition) and the $Sp(2)$
covariant Majorana condition $\lambda_a = - C \Gamma_5 \bar{\lambda}_a^T$ 
(symplectic reality condition) \cite{37}, where $C$ is the charge conjugation 
matrix. These conditions on
the 8--spinors $\lambda_a$ and $\bar{\lambda}^a$ restrict them 
to be of the form 
\begin{equation*}
\lambda_a = 
\begin{pmatrix}
i \lambda^A_{\!~~a} \\ 0 \\ 0 \\ \bar{\lambda}_{\dot{A} a} 
\end{pmatrix},
\qquad
\bar{\lambda}^a = 
( 0, - i \bar{\lambda}^{\dot{A} a}, \lambda_A^{\!~~a}, 0 ),
\qquad
C = 
\begin{pmatrix} 
\epsilon^{AB} & 0_2 \\
0_2 & \epsilon_{\dot{A} \dot{B}} 
\end{pmatrix} 
\otimes {\rm I}_2,
\end{equation*}
with the chiral and anti-chiral 2--spinors
$\lambda_A^{\!~~a}$ and $\bar{\lambda}_{\dot{A} a} =
( \lambda_A^{\!~~a} )^\dagger$, respectively. The $Sp(2)$ 
index $a$ is raised and lowered as follows, 
$\lambda_{A a} = \lambda_A^{\!~~b} \epsilon_{ba}$ and 
$\lambda_A^{\!~~a} = \epsilon^{ab} \lambda_{A b}$ where $\epsilon_{ab}$ 
is the invariant tensor of $Sp(2)$, 
$\epsilon^{ac} \epsilon_{cb} = - \delta^a_b$ (analogously for 
$\bar{\lambda}_{\dot{A} a}$).

The anti-chiral 8--spinors $\psi$ with $\psi = - i \Gamma_7 \psi$ and
$\bar{\psi} = - i \psi^\dagger \Gamma_4$ are of the form
\begin{equation*}
\psi = 
\begin{pmatrix}
0 \\ i \bar{\beta}_{\dot{A}} \\ \alpha^A \\ 0 
\end{pmatrix},
\qquad
\bar{\psi} = 
( - i \beta_A, 0, 0, \bar{\alpha}^{\dot{A}} ),
\end{equation*}
where the Weyl 2--spinors $\alpha^A$, $\bar{\beta}_{\dot{A}}$ and 
$\bar{\alpha}^{\dot{A}}$, $\beta_A$ transform according to the fundamental 
and its hermitean conjugate representation of $SL(2,C)$, respectively.

The action (B.1) is invariant under the gauge transformations
$\delta_G(\omega)$ with $\omega \equiv \omega^i T^i$,
\begin{align}
\tag{B.2}
\delta_G(\omega) A_\mu &= - D_\mu \omega,
\qquad
\delta_G(\omega) \lambda_a = [ \omega, \lambda_a ],
\qquad
\delta_G(\omega) \bar{\lambda}^a = [ \omega, \bar{\lambda}^a ],
\\
\delta_G(\omega) \psi &= \omega\,  \psi,
\qquad
\delta_G(\omega) \bar{\psi} = - \bar{\psi}\, \omega,
\qquad
\delta_G(\omega) \zeta^a = \omega\, \zeta^a,
\qquad
\delta_G(\omega) \bar{\zeta}_a = - \bar{\zeta}_a \, \omega,
\nonumber
\end{align}
and the following rigid on--shell supersymmetry transformations 
$\delta_Q = 
\bar{\rho}^{\dot{A} a} \bar{Q}_{\dot{A} a} - \rho^A_{\!~~a} Q_A^{\!~~a}$ 
with the constant chiral symplectic Majorana spinors $\rho_a$
and $\bar{\rho}^a$,
\begin{align*}
\delta_Q A_\mu & = 
\hbox{\large$\frac{1}{2}$} \bar{\rho}^a \Gamma^\mu \lambda_a -
\hbox{\large$\frac{1}{2}$} \bar{\lambda}^a \Gamma_\mu \rho_a,
\\
\delta_Q \lambda_a &
= - \hbox{\large$\frac{1}{2}$} i \Sigma^{\mu\nu} F_{\mu\nu} \rho_a +
i T^i ( \bar{\zeta}_a T^i \zeta^b ) \rho_b,
\qquad
\delta_Q \bar{\lambda}^a = \hbox{\large$\frac{1}{2}$} i \bar{\rho}^a 
\Sigma^{\mu\nu} F_{\mu\nu} -
i \bar{\rho}^b ( \bar{\zeta}_b T^i \zeta^a ) T^i,
\\
\delta_Q \zeta^a &= 2 \bar{\rho}^a \psi,
\qquad
\delta_Q \psi = i \Gamma^\mu \overset{\rightarrow}{D}_\mu \zeta^a \rho_a,
\qquad
\delta_Q \bar{\zeta}_a = 2 \bar{\psi} \rho_a,
\qquad
\delta_Q \bar{\psi} = - i \bar{\rho}^a \bar{\zeta}_a
\overset{\leftarrow}{D}_\mu \Gamma_\mu.
\end{align*}
The corresponding 8--component spinorial supercharges
\begin{equation*}
Q^a = 
\begin{pmatrix}
i Q^{A a} \\ 0_2 \\ 0_2 \\ \bar{Q}_{\dot{A}}^{\!~~a} 
\end{pmatrix},
\qquad
\bar{Q}_a = 
( 0, - i \bar{Q}^{\dot{A}}_{\!~~a}, Q_{A a}, 0 ),
\end{equation*}
obey, together with the generators $P_M$ ($M = 1, \ldots, 6$) of  
space--time translations, the $N = 2$ supersymmetry algebra,
\begin{equation}
\tag{B.3}
Q^a \otimes \bar{Q}_b + \bar{Q}_b \otimes Q^a \doteq 
- \delta^a_b ({\rm I}_8 + i \Gamma_7 )\Gamma^M ( P_M + i\delta_G(A_M) ),
\end{equation}
where the symbol $\doteq$ means that the algebra is satisfied only on--shell,
i.e., by taking into account the equations of motion. 
This algebra can also be closed off--shell at the expense of 
introducing two sets of auxiliary fields, namely the $Sp(2)$--vector field 
$\chi_{ab} = \chi_{ba}$ and the two conjugate $Sp(2)$ 
2--spinor fields $\chi_a$ and $\bar{\chi}^a \equiv \chi_a^\dagger$.

Let us now compactify the fifth and sixth dimension by a non--trivial
dimensional reduction, demanding that $A_M$ and $\lambda_a$ are independent
on $x^5$ and $x^6$ whereas $\zeta_a$ and $\psi$ are periodic in $x^5$ and
$x^6$ with the (inverse) periods $m_5$ and $m_6$, respectively,
\begin{equation*}
\partial_{5,6} A_M = 0,
\qquad
\partial_{5,6} \lambda_a = 0,
\qquad
\partial_{5,6} \zeta^a = i m_{5,6} \zeta^a,
\qquad
\partial_{5,6} \psi = i m_{5,6} \psi.
\end{equation*}
(Here it has been assumed that the higher modes are
not stimulated.) We further define
\begin{equation*}
A_5 = - i ( \phi + \bar{\phi} ),
\qquad
A_6 = \bar{\phi} - \phi,
\qquad
m_5 = m + \bar{m},
\qquad
m_6 = i ( \bar{m} - m ),
\end{equation*}
where the independence of the gauge transformations upon $x_5$ and $x_6$
have made $A_5$ and $A_6$ into a complex scalar field $\phi$, 
$\bar{\phi} \equiv \phi^\dagger$. 

After this procedure the dimensional
reduced action in four dimensions becomes
\begin{equation*}
W^{(D = 4)} = W_{\rm SYM}^{(N = 2)} + W_{\rm M}^{(Z = 2)},
\end{equation*}
with
\begin{align}
\tag{B.4}
W_{\rm SYM}^{(N = 2)} = \int d^4x\, {\rm tr} \Bigr\{&
\hbox{\large$\frac{1}{4}$} F^{\mu\nu} F_{\mu\nu} -
2 (D^\mu \bar{\phi}) (D_\mu \phi) - 
2 [ \bar{\phi}, \phi ] [ \bar{\phi}, \phi ]
\\
& - i \bar{\lambda}_{\dot{A}}^{\!~~a} (\sigma^\mu)^{\dot{A} B} 
D_\mu \lambda_{B a} + \lambda^{A a} [ \phi, \lambda_{A a} ] +
\bar{\lambda}_{\dot{A}}^{\!~~a} [ \bar{\phi}, \bar{\lambda}^{\dot{A}}_{\!~~a} ]
\nonumber
\Bigr\}
\end{align}
and
\begin{align}
W_{\rm M}^{(Z = 2)} = \int d^4x\, \Bigr\{&
i \bar{\alpha}^{\dot{A}} (\sigma^\mu)_{\dot{A} B} 
\overset{\rightarrow}{D}_\mu \alpha^B - 
2 \bar{\alpha}^{\dot{A}} ( \phi + m ) \bar{\beta}_{\dot{A}} - i (
\bar{\alpha}^{\dot{A}} \bar{\lambda}_{\dot{A} a} +
\beta_A \lambda^A_{\!~~a} ) \zeta^a
\nonumber
\\
\tag{B.5}
& + i \beta_A \overset{\leftarrow}{D}_\mu 
(\sigma^\mu)^{A \dot{B}} \bar{\beta}_{\dot{B}} - 
2 \beta_A ( \bar{\phi} + \bar{m} ) \alpha^A +
i \bar{\zeta}_a ( \lambda_A^{\!~~a} \alpha^A +
\bar{\lambda}^{\dot{A} a} \bar{\beta}_{\dot{A}} )
\\
& + \hbox{\large$\frac{1}{2}$}(D^\mu \bar{\zeta}_a) (D_\mu \zeta^a) +
\bar{\zeta}_a \{ \phi + m, \bar{\phi} + \bar{m} \} \zeta^a -
\hbox{\large$\frac{1}{4}$} 
( \bar{\zeta}_a T^i \zeta^b )( \bar{\zeta}_b T^i \zeta^a ) \Bigr\},
\nonumber
\end{align}
which is obviously invariant under the internal symmetry group $Sp(2) 
\otimes U_R(1)$ if we (formally) ascribe to $m$ and $\bar{m}$ the same 
$R$--charges as to $\phi$ and $\bar{\phi}$. The corresponding on--shell 
supersymmetry transformations in the presence of the central charges are 
\begin{align}
\phantom{\frac{1}{2}}
\delta_Q A_\mu &= \bar{\rho}^{\dot{A} a} (\sigma_\mu)_{\dot{A} B}
\lambda^B_{\!~~a} -
\bar{\lambda}^{\dot{B} a} (\sigma_\mu)_{A \dot{B}}
\rho^A_{\!~~a},
\qquad
\delta_Q \phi = i \bar{\rho}^{\dot{A} a} \bar{\lambda}_{\dot{A} a},
\qquad
\delta_Q \bar{\phi} = - i \lambda_A^{\!~~a} \rho^A_{\!~~a},
\nonumber
\\
\delta_Q \lambda_A^{\!~~a} &= - \hbox{\large$\frac{1}{2}$} i 
(\sigma^{\mu\nu})_{AB} \rho^{B a} F_{\mu\nu} +
2 i [ \bar{\phi}, \phi ] \rho_A^{\!~~a} + 
i T^i ( \bar{\zeta}_b T^i \zeta^a ) \rho_{A b} +
2 \bar{\rho}^{\dot{B} a} (\sigma^\mu)_{A \dot{B}} D_\mu \bar{\phi},
\nonumber
\\
\tag{B.6}
\delta_Q \bar{\lambda}_{\dot{A} a} &= 
\hbox{\large$\frac{1}{2}$} i
\bar{\rho}^{\dot{B}}_{\!~~a} (\sigma^{\mu\nu})_{\dot{A} \dot{B}} F_{\mu\nu} -
2 i \bar{\rho}_{\dot{A} a} [ \bar{\phi}, \phi ] - 
i \bar{\rho}_{\dot{A}}^{\!~~b} ( \bar{\zeta}_a T^i \zeta^b ) T^i +
2 (\sigma^\mu)_{\dot{A} B} \rho^B_{\!~~a} D_\mu \phi,
\end{align}
and
\begin{align}
\delta_Q \zeta^a &= 2 \bar{\rho}^{\dot{A} a} \bar{\beta}_{\dot{A}} +
2 \rho_A^{\!~~a} \alpha^A,
\nonumber
\\
\delta_Q \alpha^A &= - i \bar{\rho}_{\dot{B} a} (\sigma^\mu)^{A \dot{B}} 
\overset{\rightarrow}{D}_\mu \zeta^a + 
2 \rho^A_{\!~~a} ( \phi + m ) \zeta^a,
\nonumber
\\
\delta_Q \bar{\beta}_{\dot{A}} &= 
i \rho^B_{\!~~a} (\sigma^\mu)_{\dot{A} B}
\overset{\rightarrow}{D}_\mu \zeta^a +
2 \bar{\rho}_{\dot{A} a} ( \bar{\phi} + \bar{m} ) \zeta^a,
\nonumber
\\
\delta_Q \bar{\zeta}_a &= 2 \beta_A \rho^A_{\!~~a} +
2 \bar{\alpha}^{\dot{A}} \bar{\rho}_{\dot{A} a},
\nonumber
\\
\delta_Q \beta_A &= - i \bar{\zeta}_a 
\overset{\leftarrow}{D}_\mu (\sigma^\mu)_{A \dot{B}} \bar{\rho}^{\dot{B} a} +
2 \bar{\zeta}_a ( \phi + m ) \rho_A^{\!~~a},
\nonumber
\\
\tag{B.7}
\delta_Q \bar{\alpha}^{\dot{A}} &= 
i \bar{\zeta}_a
\overset{\leftarrow}{D}_\mu (\sigma^\mu)^{A \dot{B}} \rho_B^{\!~~a} +
2 \bar{\zeta}_a ( \bar{\phi} + \bar{m} ) \bar{\rho}^{\dot{A} a}.
\end{align}
Identifying the central charges with certain combinations
of the space--time translations on the torus, namely 
$P_5 =  Z + \bar{Z} $ and $P_6 = i (\bar{Z} - Z)$, and reverting to a 
two--spinor notation the supersymmetry algebra 
(B.3) can be recast into the 
form
\begin{align}
\{ Q_A^{\!~~a}, Q_B^{\!~~b} \} &\doteq 
- 4 \epsilon^{ab} \epsilon_{AB} ( Z + \delta_G(\phi) ),
\nonumber
\\
\tag{B.8}
\{ Q_A^{\!~~a}, \bar{Q}_{\dot{B} b} \} &\doteq  
- 2  \delta^a_b (\sigma^\mu)_{A \dot{B}} ( P_\mu + i \delta_G(A_\mu) ),
\\
\{ \bar{Q}_{\dot{A} a}, \bar{Q}_{\dot{B} b} \} &\doteq \phantom{-}
4 \epsilon_{ab} \epsilon_{\dot{A} \dot{B}} 
( \bar{Z} + \delta_G(\bar{\phi}) ),
\nonumber
\end{align}
where the central charge transformations are given by
\begin{align*}
Z V = 0,\qquad \bar{Z} V = 0
\end{align*}
 for the on-shell gauge multiplet
$V = \{ A_\mu, \lambda_A^{\!~~a}, \bar{\lambda}_{\dot{A} a}, 
\phi, \bar{\phi} \}$ and 
\begin{align*}
Z Y = m Y, 
\qquad 
Z \bar{Y} = - m \bar{Y},
\qquad
\bar{Z} Y = \bar{m} Y,
\qquad 
\bar{Z} \bar{Y} = - \bar{m} \bar{Y}
\end{align*} 
for the (massive) on-shell hypermultiplets 
$Y = \{ \alpha^A, \bar{\beta}_{\dot{A}},\zeta^a \}$ and
$\bar{Y} = \{ \bar{\alpha}^{\dot{A}}, \beta_A, \bar{\zeta}_a \}$
being hermitean conjugate to each other.

In order to derive the twisted actions (\ref{2.9}) and (\ref{2.14}) we 
identify in (B.4) and (B.5) the internal index $a$ with the 
spinor index $A$. 
It is precisely that identification which defines the twisting procedure 
\cite{2}. In addition, we introduce another set of auxiliary fields, 
$\chi_{AB} = \chi_{BA}$ and $\chi^{\dot{A}}$, $\bar{\chi}_{\dot{A}}$, in 
order to get an off--shell realization of the twisted $N = 2$ superalgebra.
This gives
\begin{align}
\tag{B.9}
W_{\rm TSYM} = \int d^4x\, {\rm tr} \Bigr\{&
\hbox{\large$\frac{1}{4}$} F^{\mu\nu} F_{\mu\nu} -
2 (D^\mu \bar{\phi}) (D_\mu \phi) - 
2 [ \bar{\phi}, \phi ] [ \bar{\phi}, \phi ] + \chi_A^{\!~~B} \chi_B^{\!~~A}
\\
& - i \bar{\lambda}_{\dot{A}}^{\!~~C} (\sigma^\mu)^{\dot{A} B} 
D_\mu \lambda_{BC} + \lambda^{AB} [ \phi, \lambda_{AB} ] +
\bar{\lambda}_{\dot{A}}^{\!~~B} [ \bar{\phi}, \bar{\lambda}^{\dot{A}}_{\!~~B} ]
\Bigr\}
\nonumber
\end{align}
and
\begin{align}
W_{\rm M} = \int d^4x\, \Bigr\{&
i \bar{\alpha}^{\dot{A}} (\sigma^\mu)_{\dot{A} B} 
\overset{\rightarrow}{D}_\mu \alpha^B - 
2 \bar{\alpha}^{\dot{A}} ( \phi + m ) \bar{\beta}_{\dot{A}} - 
i ( \bar{\alpha}^{\dot{A}} \bar{\lambda}_{\dot{A} B} +
i \beta_A \lambda^A_{\!~~B} ) \zeta^B
\nonumber
\\
\tag{B.10}
& + i \beta_A \overset{\leftarrow}{D}_\mu 
(\sigma^\mu)^{A \dot{B}} \bar{\beta}_{\dot{B}} - 
2 \beta_A ( \bar{\phi} + \bar{m} ) \alpha^A +
i \bar{\zeta}_B ( \lambda_A^{\!~~B} \alpha^A +
\bar{\lambda}^{\dot{A} B} \bar{\beta}_{\dot{A}} )
\\
& + \hbox{\large$\frac{1}{2}$} D^\mu \bar{\zeta}_A D_\mu \zeta^B +
\bar{\zeta}_A \{ \phi + m, \bar{\phi} + \bar{m} \} \zeta^A +
\bar{\zeta}_A \chi^A_{\!~~B} \zeta^B - 2 \chi^{\dot{A}} \bar{\chi}_{\dot{A}} 
\nonumber
\Bigr\},
\end{align}
with the off--shell hypermultiplets 
$Y_T = \{ \alpha^A, \bar{\beta}_{\dot{A}}, \zeta^{A}; {\bar\chi}_{\dot A} \}$ 
and
${\bar Y}_T=\{\bar{\alpha}^{\dot{A}},\beta_A,\bar{\zeta}_A;{\chi}^{\dot A}\}$.

Now we
are able to construct the complete set of $N = 2$ twisted generators,
\begin{equation*}
Q = \hbox{\large$\frac{1}{2}$} \epsilon^{AB} Q_{AB},
\qquad
\bar{Q}_\mu = \hbox{\large$\frac{1}{2}$}
i (\sigma_\mu)^{\dot{A} B} \bar{Q}_{\dot{A} B},
\qquad
Q_{\mu\nu} = \hbox{\large$\frac{1}{2}$} (\sigma_{\mu\nu})^{AB} Q_{AB},
\end{equation*}
being, respectively, a scalar $Q$, a vector $\bar{Q}_\mu$ and a self--dual 
tensor $Q_{\mu\nu}$, by substituting in (B.6) and (B.7) for $\rho^{A a}$ and
$\bar{\rho}^{\dot{A} a}$ the following expressions, 
\begin{equation*}
\rho^{AB} = \hbox{\large$\frac{1}{2}$} \rho \epsilon^{AB}, 
\qquad
\bar{\rho}^{\dot{A} B} = 
\hbox{\large$\frac{1}{2}$} i \rho^\mu (\sigma_\mu)^{\dot{A} B},
\qquad
\rho^{AB} = \hbox{\large$\frac{1}{2}$}
 \rho^{\mu\nu} (\sigma_{\mu\nu})^{AB},
\end{equation*}
where $\rho$, $\rho^\mu$ and $\rho^{\mu\nu}$ are some new global symmetry
parameters associated to $Q$, $\bar{Q}_\mu$ and $Q_{\mu\nu}$, respectively. 
Then, again disregarding the generator $Q_{\mu\nu}$, the twisted 
actions (B.9) and (B.10) will be separately invariant under the following
twisted supersymmetry transformations 
$\delta_T = \rho Q + \rho^\mu \bar{Q}_\mu$, 
\begin{align}
\delta_T A_\mu &= \bar{\rho}^{\dot{A} C} (\sigma_\mu)_{\dot{A} B}
\lambda^B_{\!~~C} -
\bar{\lambda}^{\dot{B} C} (\sigma_\mu)_{A \dot{B}}
\rho^A_{\!~~C},
\nonumber
\\
\delta_T \phi &= i \bar{\rho}^{\dot{A} B} \bar{\lambda}_{\dot{A} B},
\qquad
\delta_T \bar{\phi} = - i \lambda_A^{\!~~B} \rho^A_{\!~~B},
\nonumber
\\
\delta_T \lambda_{AC} &= - \hbox{\large$\frac{1}{2}$} i 
(\sigma^{\mu\nu})_{AB} \rho^B_{\!~~C} F_{\mu\nu} + 
2 i [ \bar{\phi}, \phi ] \rho_{AC} -
2 i \chi_C^{\!~~B} \rho_{AB} +
2 \bar{\rho}^{\dot{B}}_{\!~~C} (\sigma^\mu)_{A \dot{B}} D_\mu \bar{\phi},
\nonumber
\\
\delta_T \bar{\lambda}_{\dot{A}}^{\!~~C} &= \hbox{\large$\frac{1}{2}$} i
\bar{\rho}^{\dot{B} C} (\sigma^{\mu\nu})_{\dot{A} \dot{B}} F_{\mu\nu} -
2 i \bar{\rho}_{\dot{A}}^{\!~~C} [ \bar{\phi}, \phi ] +
2 i \bar{\rho}_{\dot{A}}^{\!~~B} \chi_B^{\!~~C} +
2 (\sigma^\mu)_{\dot{A} B} \rho^{BC} D_\mu \phi,
\nonumber
\\
\delta_T \chi_{AB} &= 
- \hbox{\large$\frac{1}{2}$} D_\mu \bar{\lambda}_{\dot{C} A}
(\sigma^\mu)^{\dot{C} D} \rho_{DB} + 
i [ \phi, \lambda^C_{\!~~A} ] \rho_{CB}
\nonumber
\\
\tag{B.11}
&\quad - \hbox{\large$\frac{1}{2}$}
 \bar{\rho}_{\dot{C} A} (\sigma^\mu)^{\dot{C} D} 
D_\mu \lambda_{DB} +
i \bar{\rho}_{\dot{C} A} [ \bar{\lambda}^{\dot{C}}_{\!~~B}, \bar{\phi} ] +
( A \leftrightarrow B )
\end{align}
and
\begin{align}
\delta_T \zeta^B &= 2 \bar{\rho}^{\dot{A} B} \bar{\beta}_{\dot{A}} +
2 \rho_A^{\!~~B} \alpha^A,
\nonumber
\\
\delta_T \alpha^A &= - i \bar{\rho}_{\dot{B} C} (\sigma^\mu)^{A \dot{B}} 
\overset{\rightarrow}{D}_\mu \zeta^C + 
2 \rho^A_{\!~~B} ( \phi + m ) \zeta^B -
2 \bar{\rho}^{\dot{B} A} \bar{\chi}_{\dot{B}},
\nonumber
\\
\delta_T \bar{\beta}_{\dot{A}} &= 
i \rho^B_{\!~~C} (\sigma^\mu)_{\dot{A} B}
\overset{\rightarrow}{D}_\mu \zeta^C +
2 \bar{\rho}_{\dot{A} B} ( \bar{\phi} + \bar{m} ) \zeta^B -
\rho^B_{\!~~B} \bar{\chi}_{\dot{A}},
\nonumber
\\
\delta_T \chi^{\dot{A}} &= - \hbox{\large$\frac{1}{2}$} i \beta_B 
\overset{\leftarrow}{D}_\mu (\sigma^\mu)^{\dot{A} B} \rho^C_{\!~~C} + 
\bar{\alpha}^{\dot{A}} ( \phi + m ) \rho^C_{\!~~C} -
\hbox{\large$\frac{1}{2}$}
 i \bar{\zeta}_B \bar{\lambda}^{\dot{A} B} \rho^C_{\!~~C}
\nonumber
\\
\tag{B.12}
&\quad + i \bar{\alpha}^{\dot{B}} \overset{\leftarrow}{D}_\mu
(\sigma^\mu)_{\dot{B} C} \bar{\rho}^{\dot{A} C} +
2 \beta_B ( \bar{\phi} + \bar{m} ) \bar{\rho}^{\dot{A} B} -
i \bar{\zeta}_C \lambda_B^{\!~~C} \bar{\rho}^{\dot{A} B},
\\
\nonumber
\delta_T \bar{\zeta}_B &= 2 \beta_A \rho^A_{\!~~B} +
2 \bar{\alpha}^{\dot{A}} \bar{\rho}_{\dot{A} B},
\nonumber
\\
\delta_T \bar{\alpha}^{\dot{A}} &= 
i \bar{\zeta}_C
\overset{\leftarrow}{D}_\mu (\sigma^\mu)^{A \dot{B}} \rho_B^{\!~~C} +
2 \bar{\zeta}_B ( \bar{\phi} + \bar{m} ) \bar{\rho}^{\dot{A} B} +
\chi^{\dot{A}} \rho_B^{\!~~B},
\nonumber
\\
\delta_T \beta_A &= - i \bar{\zeta}_C 
\overset{\leftarrow}{D}_\mu (\sigma^\mu)_{A \dot{B}} \bar{\rho}^{\dot{B} C} +
2 \bar{\zeta}_B ( \phi + m ) \rho_A^{\!~~B} +
2 \chi^{\dot{B}} \bar{\rho}_{\dot{B} A},
\nonumber
\\
\delta_T \bar{\chi}_{\dot{A}} &= 
\hbox{\large$\frac{1}{2}$} i \rho_C^{\!~~C} 
(\sigma^\mu)_{\dot{A} B} \overset{\rightarrow}{D}_\mu \alpha^B - 
\rho_C^{\!~~C} ( \phi + m ) \bar{\beta}_{\dot{A}} -
\hbox{\large$\frac{1}{2}$}
i \rho_C^{\!~~C} \bar{\lambda}_{\dot{A} B} \zeta^B
\nonumber
\\
\tag{B.13}
&\quad - i \bar{\rho}_{\dot{A} C} (\sigma^\mu)^{C \dot{B}}
\overset{\rightarrow}{D}_\mu \bar{\beta}_{\dot{B}} -
2 \bar{\rho}_{\dot{A} B} ( \bar{\phi} + \bar{m} ) \alpha^B -
i \bar{\rho}_{\dot{A} B} \lambda^B_{\!~~C} \zeta^C.
\end{align}

In order to show that the action (B.9) and (B.10) is equivalent to 
the topological action as given by (\ref{2.6}) and (\ref{2.13}) we 
carry out the following replacements:
\begin{gather*}
\chi_{AB} \rightarrow \chi_{AB} - 
\hbox{\large$\frac{1}{4}$} (\sigma^{\mu\nu})_{AB} F_{\mu\nu},
\\
\chi^{\dot{A}} \rightarrow \chi^{\dot{A}} -
\hbox{\large$\frac{1}{2}$} i \bar{\zeta}_B 
\overset{\leftarrow}{D}_\mu (\sigma^\mu)^{\dot{A} B},
\qquad
\bar{\chi}_{\dot{A}} \rightarrow \bar{\chi}_{\dot{A}} +
\hbox{\large$\frac{1}{2}$} i (\sigma^\mu)_{\dot{A} B} 
\overset{\rightarrow}{D}_\mu \zeta^B,
\end{gather*}
and we revert the spinor notation to the more familar vector notation by 
introducing the following set of fields:
\begin{alignat*}{2}
\psi_\mu &= - \hbox{\large$\frac{1}{2}$}
(\sigma_\mu)_{\dot{A} B} \bar{\lambda}^{\dot{A} B},
&\qquad
\eta &= - \hbox{\large$\frac{1}{2}$} i \epsilon_{AB} \lambda^{AB},
\\
\chi_{\mu\nu} &= - \hbox{\large$\frac{1}{2}$}
i (\sigma_{\mu\nu})_{AB} \lambda^{AB},
&\qquad
\lambda_{\mu\nu} &= \hbox{\large$\frac{1}{2}$}
(\sigma_{\mu\nu})_{AB} \chi^{AB}.
\end{alignat*}
Then, one easily verifies that the resulting action is just the topological
action (\ref{2.6}), (\ref{2.13}) for $\xi = 1$ with the Pontrjagin term
subtracted, i.e., it is determined by the invariant polynomial 
${\rm tr}\, \phi^2$ through the very attractive form \cite{21},
\begin{equation*}
W_{\rm TSYM} = W_{\rm T} - \frac{1}{4} \int d^4x\, {\rm tr}\,
F^{\mu\nu} \tilde{F}_{\mu\nu} = - \frac{1}{24} \epsilon^{\mu\nu\rho\sigma} 
\bar{Q}_\mu \bar{Q}_\nu \bar{Q}_\rho \bar{Q}_\sigma 
\int d^4x\, {\rm tr}\, \phi^2,
\end{equation*}
thus also giving a suggestive idea of the usefulness of the vector operator 
$\bar{Q}_\mu$. Performing the same replacements in (B.10) we arrive at the 
matter action (\ref{2.14}) for $\xi = 1$.

Finally, in order to establish the relationship between the twisted $N = 2$ 
SYM and TYM one identifies the $R$--charge with the ghost number, 
i.e., one goes over from a conventional QFT to a cohomological one \cite{4,5}. 
This is achieved by simply setting in (B.11) -- (B.13) the ghost 
$\rho$ associated to the singlet operator $Q$ equal to one, i.e., $\rho = 1$.
Thereby, the ghost numbers of the remaining fields have to be redefined.
In this way one recovers the topological shift symmetry and the vector 
supersymmetry introduced in (\ref{2.7}), (\ref{2.8}), (\ref{2.15}) 
and (\ref{2.17}) for $\xi = 1$.


\end{document}